\documentclass[sigplan,twocolumn]{acmart}
\acmSubmissionID{<17>}
\renewcommand\footnotetextcopyrightpermission[1]{}
\usepackage{enumitem}
\usepackage{wrapfig}
\usepackage{subfig}
\usepackage[linesnumbered,ruled,vlined]{algorithm2e}
\usepackage[noend]{algpseudocode}

\AtBeginDocument{%
  }

\acmISBN{978-1-4503-XXXX-X/2018/06}

\begin{document}

\title{CoLLM: Continuous Adaptation for SLO-Aware LLM Serving on Shared GPU Clusters}

\author{Shaoyuan Huang}
\affiliation{%
  \institution{Tianjin University}
  \city{Tianjin}
  \country{China}}
\email{hsy_23@tju.edu.cn}

\author{Yunfeng Zhao}
\affiliation{%
  \institution{Tianjin University}
  \city{Tianjin}
  \country{China}}
\email{yfzhao97@tju.edu.cn}

\author{Na Yan}
\affiliation{%
  \institution{King's College London}
  \city{London}
  \country{United Kingdom}}
\email{na.2.yan@kcl.ac.uk}

\author{Tiancheng Zhang}
\affiliation{%
  \institution{Tianjin University}
  \city{Tianjin}
  \country{China}}
\email{tianchengzhang@tju.edu.cn}

\author{Xiaokai Wang}
\affiliation{%
  \institution{Tianjin University}
  \city{Tianjin}
  \country{China}}
\email{xiaokaiwang@tju.edu.cn}

\author{Xiaofei Wang}
\affiliation{%
  \institution{Tianjin University}
  \city{Tianjin}
  \country{China}}
\email{xiaofeiwang@tju.edu.cn}

\author{Wenyu Wang}
\affiliation{%
  \institution{Paiou Cloud Computing (Shanghai) Company, Ltd.}
  \city{Shanghai}
  \country{China}}
\email{wayne@pplabs.org}

\author{Yansha Deng}
\affiliation{%
  \institution{King's College London}
  \city{London}
  \country{United Kingdom}}
\email{yansha.deng@kcl.ac.uk}

\renewcommand{\shortauthors}{Trovato et al.}

\begin{abstract}
% 一句话介绍场景(deep learning (DL) clusters),效果(improves scheduling performance)，核心方法(adaptively co-optimizing inter-dependent factors both at the per-job level and at the cluster-wide level).
% Pollux improves scheduling performance in deep learning (DL) clusters by adaptively co-optimizing inter-dependent factors both at the per-job level and at the cluster-wide level.
LLM serving systems often suffer from GPU resource underutilization due to highly dynamic workloads and GPU over-provisioning. While existing works consider scheduling additional training workloads through temporal or spatial GPU sharing to enhance utilization, they treat training and inference as disjoint tasks, leading to costly model redeployment with minimal improvements in service quality.
% The widely adopted Parametric Efficient Fine Tuning (PEFT) technique not only promises the rise of personalized model services, but also offers a practical opportunity for model sharing between inference and fine-tuning tasks through a unified interface, supporting quality-aware serving.
% The fundamental concept behind CoLLM 源于我们对popular Parametric Efficient Fine Tuning (PEFT) technique的洞察，其不仅promises the rise of personalized model services, but also offers a practical opportunity for model sharing between inference and fine-tuning tasks through a unified interface, supporting quality-aware serving.
In this paper, we introduce CoLLM, the first co-orchestrating system for joint Low-Rank Adaptation (LoRA) based LLM inference and fine-tuning.
% The core idea of CoLLM stems from our insight that the Parameter-Efficient Fine-Tuning (PEFT) technique, widely adopted for personalized model adaptation, also offers a unified interface for model sharing between inference and training tasks.
The core idea behind CoLLM is that the Parameter-Efficient Fine-Tuning (PEFT) offers a practical opportunity for model sharing between inference and training through a unified PEFT interface.
Building on this, CoLLM enables training–inference co-execution on the same replicas with shared model weights, leveraging idle GPU resources and feeding fine-tuning gains into real-time inference.
% CoLLM defines replica behavior via a novel states management mechanism and incorporates three key components: inter-replica orchestrating for launching opportunistic fine-tuning tasks, intra-replica coordination with interference-aware batch optimization, and subflow-based dispatching to regulate fluctuating request streams.
CoLLM defines replica behavior via a novel state management mechanism and incorporates three key components: inter-replica orchestration that opportunistically launches fine-tuning tasks on idle replicas, intra-replica coordination that jointly optimizes training and inference batch sizes with concurrent interference-aware latency modeling, and subflow-based dispatching that transforms unpredictable request streams into controlled subflows tailored to each replica.
Evaluations on production-scale LLM serving traces and GPU testbeds show that CoLLM simultaneously improves inference throughput and response quality. Compared to state-of-the-art serving systems, CoLLM achieves up to 1.5× higher goodput (throughput that meets latency constraints) and 2.2× better quality weighted goodput, with GPU utilization improved by 40\%.
\end{abstract}

\maketitle

\section{Introduction}
% 从相关文献的数量可以看出，从训练服务到推理服务的关注也是一个重要趋势，尤其是在大模型应用越来越普遍的背景下。

Large language models (LLMs) have become foundational to modern AI inference services, powering applications such as chatbots, coding assistants, and autonomous agents~\cite{openai, codellama, jiang2024surveylargelanguagemodels, deepseekai2024deepseekv3technicalreport}.
To make LLMs adaptable to different tasks and users, recent work has proposed parameter-efficient fine-tuning (PEFT), such as Low-Rank Adaptation (LoRA) \cite{hu2022lora, 2023qlora}, which offers a lightweight mechanism to personalize LLMs without modifying their base weights. This technique promises the rise of multi-tenant LoRA LLM serving systems, offering "LoRA-as-a-Service" capabilities~\cite{Wu2024, Punica, SLoRA}. 
% These systems enable efficient personalization of large pre-trained models for diverse domain-specific applications, leveraging PEFT with localized examples. Such examples are often generated as user-issued requests in multi-tenant systems—e.g., domain-specific prompts, feedback-corrected responses, or synthesized queries that are logged in real time.
These systems serve domain-specific applications through personalized adapters, which are fine-tuned via PEFT on tenant-provided data or user-issued requests, such as domain-specific prompts and feedback-corrected responses.
% that are logged in real time.

To ensure latency Service Level Objectives (SLOs)~\cite{201468,273804, seer, qmrgnn} and stable throughput, LLM serving systems often over-provision GPU resources to deploy model replicas. However, such redundant provisioning leads to significant underutilization, particularly during off-peak periods, where request arrival rates can drop by an order of magnitude. 
% Efficient utilization of expensive GPU resources while maximizing serving throughput is critical for current LLM serving systems. However, to ensure strict latency Service Level Objectives (SLOs)~\cite{201468,273804}, production deployments often over-provision GPU resources to deploy model replicas, leading to significant GPU underutilization, especially during off-peak periods when request arrival rates can drop by more than an order of magnitude.
To improve GPU utilization, prior work has explored GPU multiplexing strategies. \textit{Temporal sharing}~\cite{Chen2023} opportunistically schedules additional workloads during idle periods. As training task is typically compute-intensive and latency-tolerant, it serves as a widely adopted filler workload~\cite{flexmm, mudi2024}. However, training jobs are costly to preempt and slow to adapt \cite{227623, euro_preempt}, which can cause delayed inference when load surges or trigger expensive request migration, such as~\cite{Wu2024, Zhang2023}. These issues fundamentally limit the effectiveness of pure temporal sharing. 
\textit{Spatial sharing}~\cite{acm_spatial_sharing1, yu2022survey} offers an alternative by enabling concurrent inference and training execution on the same device via GPUs multi-stream or Multi-Process Service (MPS)~\cite{nvidia_mps}, which improves utilization with minimal serving performance degradation. However, most existing spatial sharing systems~\cite{mudi2024, Luo2024} target conventional Deep Learning (DL) models and require deploying separate replica instances for each task, even when sharing the same backbone. In LLM-scale workloads, such redundancy is very costly in terms of memory and cold-start latency. 

More importantly, current approaches focus solely on resource sharing, leaving inference agnostic to training progress and unable to benefit from fine-tuning gains. Such isolation overlooks a critical opportunity of \textit{quality-aware serving through model sharing}, where serving directly benefits from ongoing fine-tuning by sharing model weights, enabling real-time improvements in response quality. As LoRA-based LLM systems also require domain-specific model quality such as freshness, personalization, and domain adaptation~\cite{Wu2024, 2024codellama}, there is a growing need for a unified design where inference and fine-tuning share model weights through a unified PEFT interface to allow real-time quality adaptation.

Additionally, adaptive batching is another widely adopted strategy in LLM serving systems, which enables replicas to process requests in groups with dynamically adjusted batch sizes to improve GPU utilization and throughput~\cite{Qiao2021, Zhang2023, mudi2024}. 
However, its effectiveness is limited under concurrent task interference, as most batching strategies rely on a stable linear relationship between inference latency and batch size~\cite{Nexus2019, Wu2024}, yet real-world latency characteristics are highly sensitive to resource utilization and interference from co-running workloads. In the presence of concurrent fine-tuning, inference latency exhibits dynamic shifts, violating basic linear assumptions and complicating batching decisions.

Last but not least, inference workload unpredictability further challenges system stability. 
% Prior studies such as Mudi~\cite{mudi2024} and Shepherd~\cite{Zhang2023} demonstrate that sub-second request arrival rates exhibit high variance, making short-term load patterns unpredictable. 
In real-world system, sub-second request arrival rates often exhibit high variance, making short-term load patterns unpredictable~\cite{mudi2024, Zhang2023}. 
However, existing systems primarily adopt reactive strategies, such as resource reallocation~\cite{Zhang2023, Wu2024} to mitigate performance degradation caused by fluctuating workloads, rather than proactively regulating the request stream. Such lagged reaction undermines static dispatching and destabilizes the adaptive batching optimizations under tight SLO constraints.

% These limitations highlight the need for a unified system that co-optimizes inference serving and fine-tuning, while preserving low latency, high throughput, and personalized model quality under dynamic conditions.

To address these challenges, we present \textbf{CoLLM}, the first \textbf{C}o-orchestrating system for joint \textbf{L}oRA LLM \textbf{I}nference and \textbf{F}ine-tuning that enables quality-aware services in GPU clusters. CoLLM enables concurrent LoRA inference and fine-tuning on shared replicas by combining temporal and spatial GPU multiplexing with model-level sharing. This design fully leverages the complementary nature of the two workloads, utilizing idle resources for fine-tuning while feeding model quality improvements into real-time inference. CoLLM integrates three tailored strategies to achieve this goal: dynamic replica co-orchestration, interference-aware batch coordination, and subflow-based request dispatching. In particular:

% orchestrating replica behaviors, coordinating batching under concurrent tasks, and stabilizing request streams under unpredictable workloads.

% CoLLM dynamically exploits the complementarities between LoRA inference and fine-tuning at both the resource and model levels, preserving high throughput, and personalized model quality under dynamic workloads. It achieves this through three tailored strategies that orchestrate replica behavior, coordinate batching with interference, and stabilize request stream under unpredictable workloads.

\textit{Inter-replica Co-orchestration.} CoLLM introduces a novel repli-ca state management mechanism to govern replica behavior. Through the \textbf{Fine-tune Task Launcher}~(\S\ref{sec:FTL}), CoLLM opportunistically orchestrates fine-tuning tasks on idle replicas via temporal multiplexing, and dynamically supports early stopping to release resources during load surges. The federated learning (FL) scheme further aggregates collaborative fine-tuning processes across distributed replicas, improving both resource utilization and model quality.

\textit{Intra-replica coordination.} To maximize intra-replica efficiency, CoLLM applies spatial multiplexing to support concurrent inference and fine-tuning on the same model instance. The \textbf{Inference-Training Coordinator}~(\S\ref{sec:ITC}) builds interference-aware latency models and optimizes per-replica batch sizes to maximize training goodput while opportunistically serving inference workload under SLO constraints.

\textit{Subflow-based Request Dispatching.} To address workload burstiness and unpredictability, CoLLM introduces a subflow-based \textbf{Dispatcher}~(\S\ref{sec:dispatcher}) that regulates the request stream into paced subflows per replica. A two-phase adjustment strategy is incorporated to ensure the long-term robustness of latency modeling while maintaining short-term responsiveness to request burstiness and fluctuations in model quality.

% A two-phase adjustment mechanism combines long-term latency modeling and short-term priority-based tuning, achieving high-throughput and quality-aware inference serving.

Through these designs, CoLLM significantly improves GPU utilization, enables real-time model quality enhancement, high-goodput and quality-aware inference under dynamic workloads. We implement and evaluate CoLLM on a 32$\times$ H20 GPUs testbed with production-scale LLM serving workloads. Compared to state-of-the-art LLM serving systems, CoLLM improves inference goodput by up to 1.5× and response quality weighted goodput by up to 2.2×, maintains over 70\% GPU utilization under low workload. It also scales efficiently with workload and incurs minimal overhead.

\section{Background and Motivation}\label{sec:motivation}
% 1. 在推理服务阶段，请求的周期性导致的资源冗余分配会造成资源不充分利用。
% 2. 在微调训练阶段，不同client的资源和数据异构性会导致微调（FL PEFT）对资源的利用不充分。
% 3. Fine-tuning总会提升模型performance，从而提升serving quality，但serving与fine-tuning的隔离导致提升难以及时体现。
% 4. GPU sharing的方式与时延SLO的存在决定了请求服务的高优先级（即必须保证请求的等待+处理时间<SLO），需要由request dispatcher+coordinator来保证（不能将request分发给无法执行的replica，必须优先服务即将超时的request）。
% 要探究资源-时间消耗-请求服务-fine-tuning之间的建模关系。

\subsection{Joint LoRA LLM Inference and Fine-tuning}\label{sec:model_sharing}

\begin{figure}[tbp]
    \centering
    \captionsetup{skip=3pt}
    \includegraphics[width=\columnwidth]{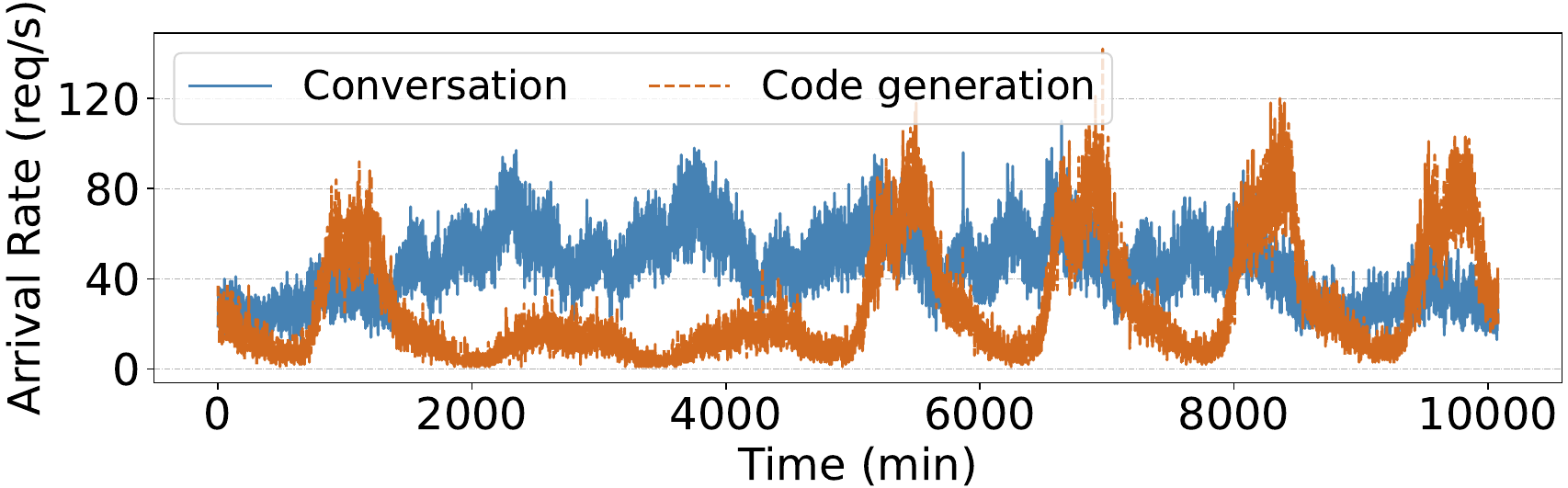}
    \caption{Arrival rate distribution.}
    \label{fig:motivation11}
    \vspace{-0.5cm}
\end{figure}

Despite the high cost of GPUs and the expertise of operators, LLM inference serving systems often suffer from low GPU utilization. A recent NVIDIA GTC 2025 report~\cite{nvidia_report} notes that current GPU utilization in AI facilities hovers around 30\% on average, and the central cause of this underutilization is GPU over-provisioning and highly fluctuating workloads~\cite{Wu2024, mudi2024}.

% Weights\& Biases also reported that nearly one-third of users have an average GPU utilization of less than 15\%~\cite{}.

To illustrate the dynamics of online LLM inference workloads, we analyze two LLM inference service traces from Azure~\cite{Azure_Dataset}. As depicted in Fig. \ref{fig:motivation11}, the request arrival rate (req/s) during low-traffic periods can drop to less than 0.7\% of the peak rate, while surges can lead to a 440\% sudden increase in requests. Such unpredictable workload fluctuations would inevitably result in underutilization of over-provisioned resources during low-demand periods.

% 这种低利用率催生了对GPU资源的时空复用技术，例如工作\cite{Deepboot}就探索了将空闲的GPU分配给训练任务并在负载提高时及时收回的调度框架，Mudi和Luo2024则efficiently multiplexing DL inference services with training tasks through spatial sharing. 

\begin{figure}[tbp]
    \centering
    \captionsetup{skip=4pt}
    \includegraphics[width=0.9\columnwidth]{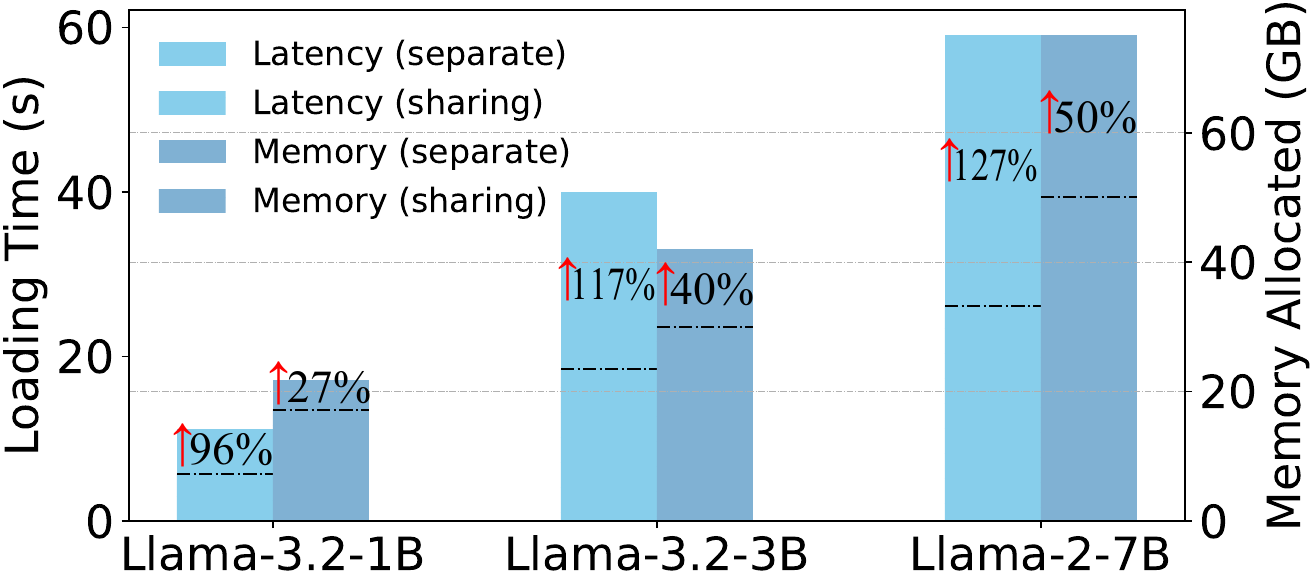}
    \caption{Additional overhead caused by separate replica loading of different models on one Nvidia A100 80G GPU.}
    \label{fig:motivation12}
    \vspace{-0.5cm}
\end{figure}

This low utilization has spurred the development of temporal and spatial GPU multiplexing~\cite{280768, mudi2024, Chen2023, Luo2024}. For example, DeepBoot~\cite{Chen2023} explores scheduling frameworks that allocates idle GPUs to training tasks and reclaims them as load increases. Mudi~\cite{mudi2024} and \cite{Luo2024} efficiently multiplex DL inference services with training tasks through spatial sharing. However, most of these systems target conventional DL models and require fully reloading the replica for concurrent execution—even for the same model. In large-scale LLMs, such redundancy introduces significant overhead. Fig.~\ref{fig:motivation12} compares the replica loading time and run-time memory footprint of the separate loading and model sharing approach for concurrent fine-tuning and inference tasks under the same workload. It can be seen that reloading a 7B-scale LLM can additionally result in more than 30s (127\%) of latency and more than 20G (50\%) of GPU memory usage, and this extra load percentage tends to increase with model size.

\begin{figure}[tbp]
    \centering
    \captionsetup{skip=3pt}
    \includegraphics[width=\columnwidth]{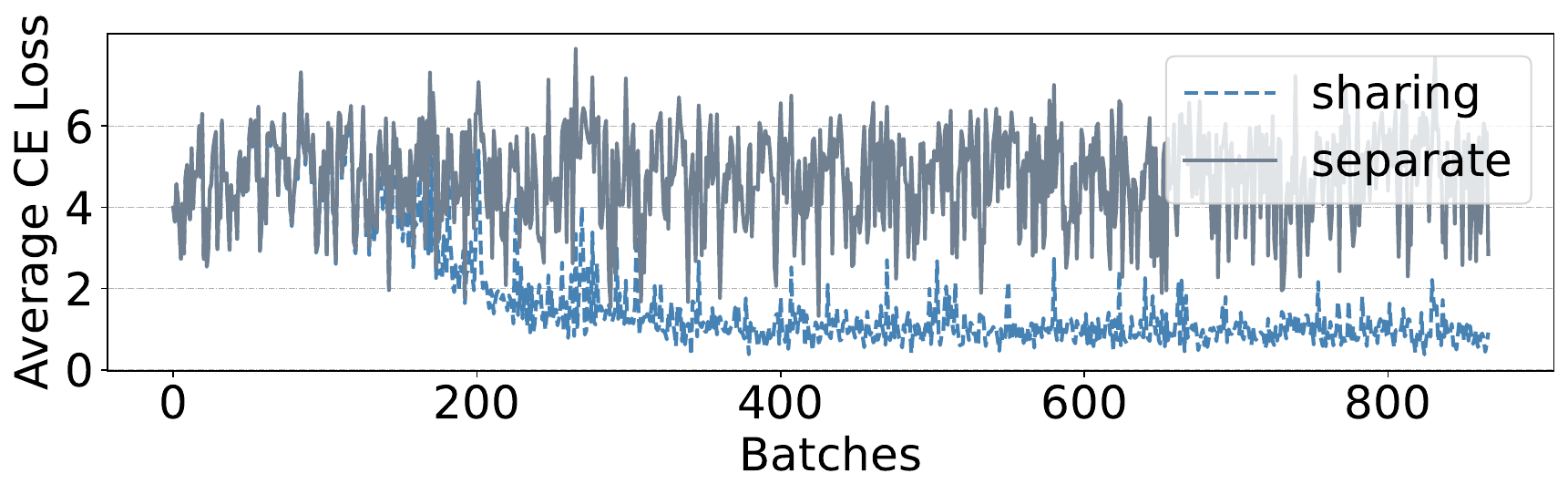}
    \caption{CE Loss of the model serving under separate loading and model sharing.}
    \label{fig:motivation13}
    \vspace{-0.5cm}
\end{figure}

\textbf{CoLLM's model sharing.} CoLLM enables \textit{model sharing} by concurrently running inference and training on shared replica and model instance. Since LoRA-based PEFT provide a unified interface for inference and fine-tuning by injecting low-rank adapters, making it possible to share model weights and intermediate activations. CoLLM loads LLM into GPU memory once and spawns two isolated Python subprocesses, one for inference and one for fine-tuning, via multiprocessing. Each subprocess maintains a separate CUDA stream and computation graph. The fine-tuning process retains its own gradients and optimizer state, while inference operates on updated model snapshots and cached KV tokens (other technical details are provided in \S\ref{sec:implementation} ). 
% This design ensures compute concurrency and memory safety, enabling CoLLM to interleave fine-tuning and inference on a single model without duplication or mutual interference.

% model sharing的另一个非常显著的收益是实时的推理质量提升，如图1b所示，我们评估了Llama-2-7B在同一批code generation请求下的推理回复质量（evaluated via token-level Cross-Entropy Loss (CE Loss), a standard metric that quantifies how well the model’s predicted tokens align with the ground-truth responses in supervised instruction data）。可以看到，随着并发的fine-tuning任务的进行，CE Loss有着明显的下降趋势，这说明推理任务也实时享受到了模型的质量提升。

Another notable benefit of model sharing is the enhancement of real-time inference quality. As illustrated in Fig.~\ref{fig:motivation13}, we evaluate the quality\footnote{Inference quality is measured using token-level cross-entropy (CE) loss, a standard metric that quantifies the alignment between the model’s predicted tokens and the ground-truth responses from supervised instruction data.} of responses generated by a LLaMA-2-7B model on a fixed workload of code generation queries. The results show a clear downward trend in CE loss as fine-tuning progresses concurrently with inference under a model sharing setup, indicating that inference tasks benefit immediately from ongoing model updates and experience continuous quality improvement.

\subsection{Latency Modeling and Throughput Optimization}\label{sec:motivation2}
Service latency for batch requests is often assumed to follow a simple linear model with respect to batch size~\cite{Nexus2019, Zhang2023, Qiao2021}, typically modeled as $y = \alpha x + \beta$. Here, $x$ is the batch size and $y$ is the corresponding latency. This linearity underpins the common practice of increasing batch size to improve inference throughput $x/y$, as larger batches help amortize the fixed cost $\beta$ and saturate compute resources~\cite{Zhang2023}.

\begin{figure}[tbp]
\centering
\subfloat[$\alpha$ floating with avg infer batch size]{\label{fig:alpha_floating}
\includegraphics[width=0.48\columnwidth]{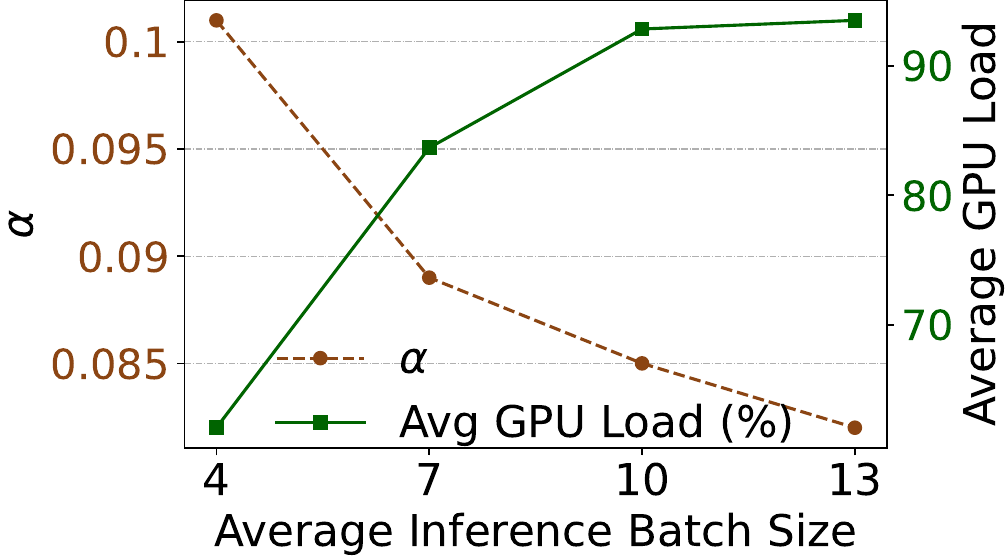}
}
\subfloat[Latency under interference]{\label{fig:latency_modeling}
\includegraphics[width=0.481\columnwidth]{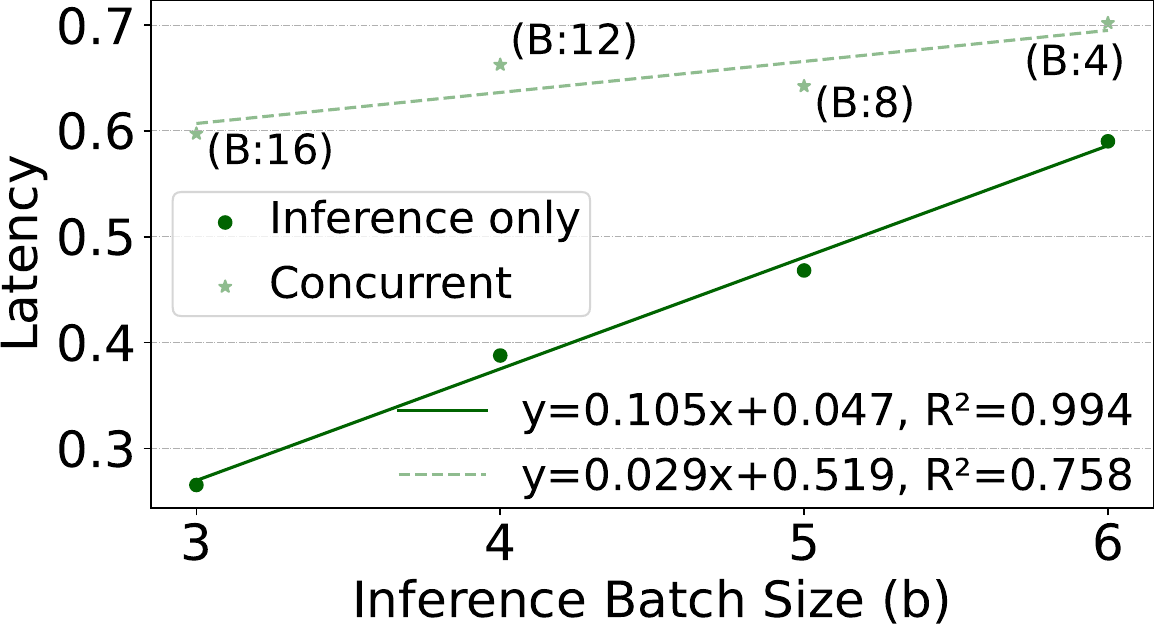}
}
\caption{Instability and interference sensitivity of latency modeling}
\vspace{-0.5cm}
\end{figure}

In practice, however, this modeling assumption is fragile. First, excessive batch sizes may violate latency SLOs, introducing a trade-off between throughput and tail-latency compliance. Second, even for a fixed model and device, model parameters $\alpha$ and $\beta$ are not static. As shown in Fig.~\ref{fig:alpha_floating}, for Llama-2.1-7B on an A100 GPU, $\alpha$ varies significantly as average infer batch size and GPU utilization increases, and only approaches stability at saturation. Unfortunately, the batch size required to reach saturation often exceeds the SLO-compliant upper bound, and bursty traffic rarely provides sufficient inflow to support such large batches.

% Moreover, GPU execution control mechanisms offer limited support for fixing resource usage. NVIDIA MPS, for instance, allows resource sharing but cannot enforce strict caps on compute usage. When resources are idle, processes may exceed their nominal limits, making it difficult to consistently execute at a target utilization ratio. This variability challenges static latency models, which must generalize across fluctuating usage levels and thus require diverse input profiles and frequent online re-fitting.

% 当我们在同一设备上为batch size=[3,4,5,6]的推理任务并发上batch size=[16,12,8,4]的训练任务后，不同infer batch size的推理时延之间的差距变得不再那么明显。
Moreover, in the presence of concurrent fine-tuning, latency modeling becomes further complicated. Co-located inference and training tasks contend for shared resources, including compute, memory bandwidth, and PCIe lanes, resulting in cross-task interference. As illustrated in Fig.~\ref{fig:latency_modeling}, when fine-tuning task with batch size\footnote{We apply gradient accumulation with a factor of 8, so the effective training batch size per update is 8× larger than the per-step batch size.}=(16,12,8,4) is co-located with the inference task with batch size=(3,4,5,6), the gap between the inference latencies of different infer batch sizes becomes less pronounced. The corresponding latency model’s $R^2$ score drops from 0.994 to 0.758, indicating degraded predictability. A similar impact is observed for training-time modeling under interference. These findings motivate the need for interference-aware latency modeling that accounts for both batch size and co-running task intensity.

\subsection{Inference Workload Unpredictability}\label{motivation3}
% As demonstrated in \cite{Zhang2023}, the average request arrival rates can be quite unpredictable at smaller time granularities (e.g., milliseconds) that must be considered to meet per-request SLO deadlines. %该工作研究了不同的时间窗口下请求到达数的coefficient of variance，发现 the high coefficient of variance at even minute-granularity makes sub-second request arrival patterns nearly impossible to predict. 

% 这种不可预测性会从根本上限制服务系统的资源利用率与服务SLO保证，并使得之前所探讨的针对throughput的最优化决策效果欠佳甚至失效。To explain why, 我们定义\textit{Ideal Serving Mode}, 并观察在给定的请求负载下，使用轮询分发，一种不加任何优化，直接将请求分发到装载了所需model的replica的分发方式，与\textit{Ideal Serving Mode}之间的性能差距。

Inference request arrivals in real-world systems are highly bursty and exhibit substantial short-term variability. As dem-onstrated in~\cite{Zhang2023}, even minute-level request traces show large coefficients of variation, making it impossible to accurately predict request arrivals at millisecond-level granularity—an essential timescale for per-request SLO guarantees. Such unpredictability not only limits resource utilization, but also undermines the effectiveness of batching and dispatch decisions designed to maximize throughput.

% In this pace, each batch completes exactly as the next set of requests arrives at the replica, resulting in a saturated yet SLO-compliant serving throughput. 
To illustrate the challenge, we define the \textit{Ideal Serving Mode} as the optimal serving pace in which each replica continuously processes inference requests at the maximum batch size that meets the latency SLO. Formally, let:
\begin{itemize}
    \item $b_i$ denotes the inference batch size on replica $i$,
    \item $t_i(b_i)$ is the processing time for a batch of size $b_i$,
    \item $\tau$ denotes the latency SLO (deadline),
    \item $\lambda_i$ denotes the average request arrival rate to replica $i$ (requests per second).
\end{itemize}

We define that replica $i$ operates in \textit{Ideal Serving Mode} if and only if the following conditions hold: (1) maximum batch size within latency budget: $t_i(b_i^\star) = \tau$ and (2) exact request arrival to fill next batch: $b_i^\star = \lambda_i \cdot \tau$. That is, the replica processes batches that fully utilize the latency budget $\tau$, and receives exactly $b_i^\star$ requests during each processing cycle—eliminating any queueing latency and achieving SLO-saturated throughput.

%(一些实验结果等待我来补充)

% 可以观察到，即使是使用了最优的infer batch size，轮询分发的GPU资源利用率和服务goodput都远低于理想水平，当负载较低，replica所需的infer batch size得不到满足，导致资源利用率欠佳；而负载较高时，又会因为replica内较高的排队时延，导致未被SLO，服务失败。不论是轮询还是基于时延的贪心分发，这类面向单个请求的分发策略由于无法主动克制负载的不可预测性，使得请求流自身的不稳定性被转嫁到服务replica上，导致原有的最优化决策失效和次优的系统表现。

% However, common dispatch strategies such as round-robin or greedy latency-aware dispatching~\cite{Wu2024} fail to maintain this mode.为了说明这一结论，我们对比了replica在相同的总请求量与最优batch size $b_i^\star$设置下，以ideal service mode分发请求与按照round-robin以request arrival rate不加控制地分发，GPU的计算资源利用率与实时请求成功服务数（under latcny SLO）。 As Fig.~\ref{fig:motivation3} shown，相较于ideal service mode，round-robin使得the stochastic nature of inference workload propagates instability into LLM serving, invalidating throughput-optimized batching strategies and yielding suboptimal system performance. When arrival rates are low, batch sizes fail to saturate the GPU, resulting in poor utilization (Fig.~\ref{fig:gpu_load}). When arrival rates spike, replicas experience excessive queueing delays, leading to SLO violations (Fig.~\ref{fig:request_served}). 

\begin{figure}[tbp]
\centering
\captionsetup{skip=3pt}
\subfloat[GPU SM utilization]{\label{fig:gpu_load}
\includegraphics[width=0.9\columnwidth]{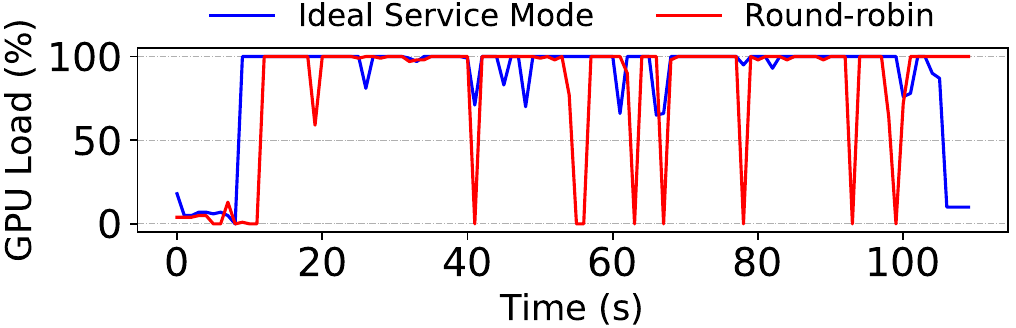}
}
\\
\vspace{-2mm}
\subfloat[Requests served that meet SLOs]{\label{fig:request_served}
\includegraphics[width=0.9\columnwidth]{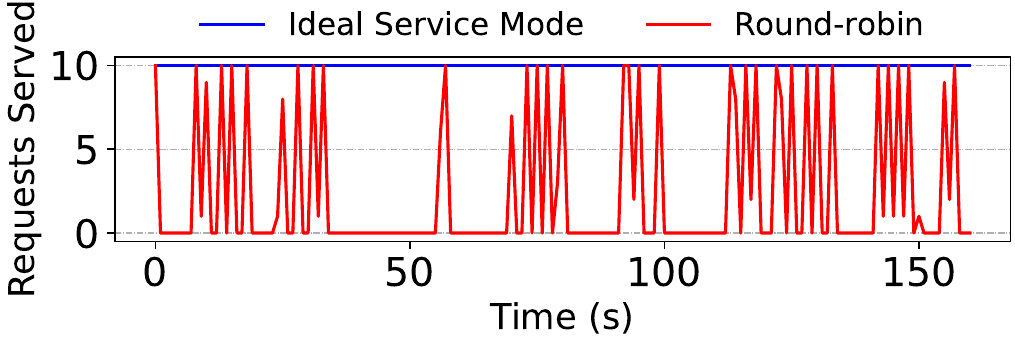}
}
\caption{Fluctuating workloads lead to significant deviations from the ideal service model.}
\label{fig:motivation3}
\vspace{-0.5cm}
\end{figure}

However, common dispatching strategies such as round-robin or greedy latency-aware schemes~\cite{Wu2024} fail to sustain this mode due to the lack of proactive control over the request stream. To illustrate, we compare an ideal-mode dispatcher against a round-robin dispatcher, both under the same total load and optimal per-replica batch size $b_i^\star$. As shown in Fig.~\ref{fig:motivation3}, compared to the ideal-mode dispatcher, round-robin dispatching allows the stochastic nature of inference workloads to propagate directly to the replicas, destabilizing throughput-optimized batching and degrading overall performance. When arrival rates are low, batches fail to saturate GPU capacity, leading to poor utilization (Fig.~\ref{fig:gpu_load}); when arrival rates spike, queueing latency accumulate, resulting in SLO violations (Fig.~\ref{fig:request_served}).
% even with the same workload and optimal inference batch size, round-robin dispatching greatly underperforms the ideal service model.
This motivates the need for a dispatching strategy that explicitly mitigates short-term workload variability, stabilizing per-replica execution and enabling latency-aware optimizations to take effect.

\section{CoLLM Overview}
% 其解决了一个双向切本质互补的问题，即如何在单纯的推理系统中无缝融入模型微调，并实时地利用微调后的模型提供推理服务，以提升对有限GPU的资源利用率，提高模型服务性能，同时保证时延的SLO。CoLLM的核心在于一套贯穿全局的replica状态管理机制，以及用于启动定制化微调过程的fine-tune task launcher和维护训练-推理复用的inference-training coordinator。除此之外，一般推理系统中所具备的replica allocator和request dispatcher也随着状态机制的引入进行了适应性调整。
CoLLM is a joint inference serving and fine-tuning system designed to serve multiple LoRA-based LLM models in a cluster. At the cluster level, CoLLM addresses the complementary problem of seamlessly integrating fine-tuning into a predominantly inference-oriented system and leveraging fine-tuned models to improve service quality. 
% by state-driven replica behavior switching
% In the replica level, CoLLM精心为处在不同状态的replica定制任务内部参数配置，并自适应地分发请求。二者的交叉使得CoLLM圆满地实现了以下多个目标：
At the replica level, CoLLM meticulously tailors the task-internal configurations for replicas. The intersection of the two-level optimization enables CoLLM to satisfactorily achieve multiple system goals: 1) enhancing the GPU resources utilization, 2) maximizing inference service throughput with better model quality, and 3) ensuring compliance with latency SLOs. 

\begin{figure}[tbp]
    \centering
    \includegraphics[width=\columnwidth]{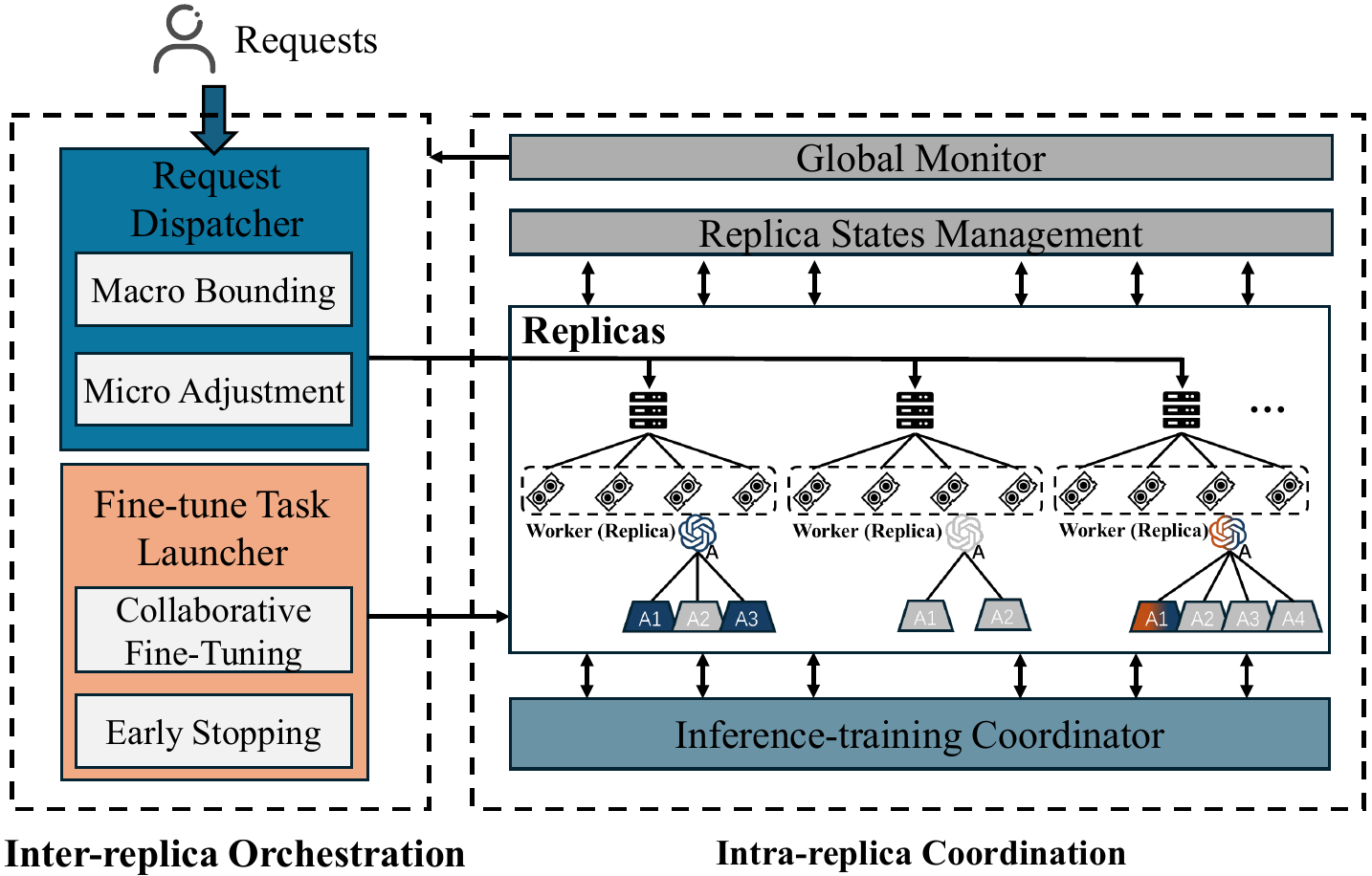}
    \caption{CoLLM overview.}
    \label{fig:CoLLM}
    \vspace{-0.5cm}
\end{figure}

As shown in Fig. \ref{fig:CoLLM}, CoLLM features a novel global replica state management mechanism, a \textbf{Fine-tune Task Launcher} (\S\ref{sec:FTL}) for orchestrating customized fine-tuning processes, and an \textbf{Inference-training Coordinator} (\S\ref{sec:ITC}) to coordinate complementarities and interference between concurrent tasks. Additionally, CoLLM introduces a customized \textbf{Request Dispatcher} (\S\ref{sec:dispatcher}) to manage chaotic unpredictable loads, ensuring that CoLLM's optimization decisions are effective in any complex system-load context.

% 客制化了传统的请求分发器，使CoLLM最大程度掌控混沌（不可预知）的负载，保证CoLLM的最优化决策在如此复杂的系统任务环境下的效果。

\begin{figure*}[htbp]
    \vspace{0.45cm}
    \centering
    \includegraphics[width=0.85\linewidth]{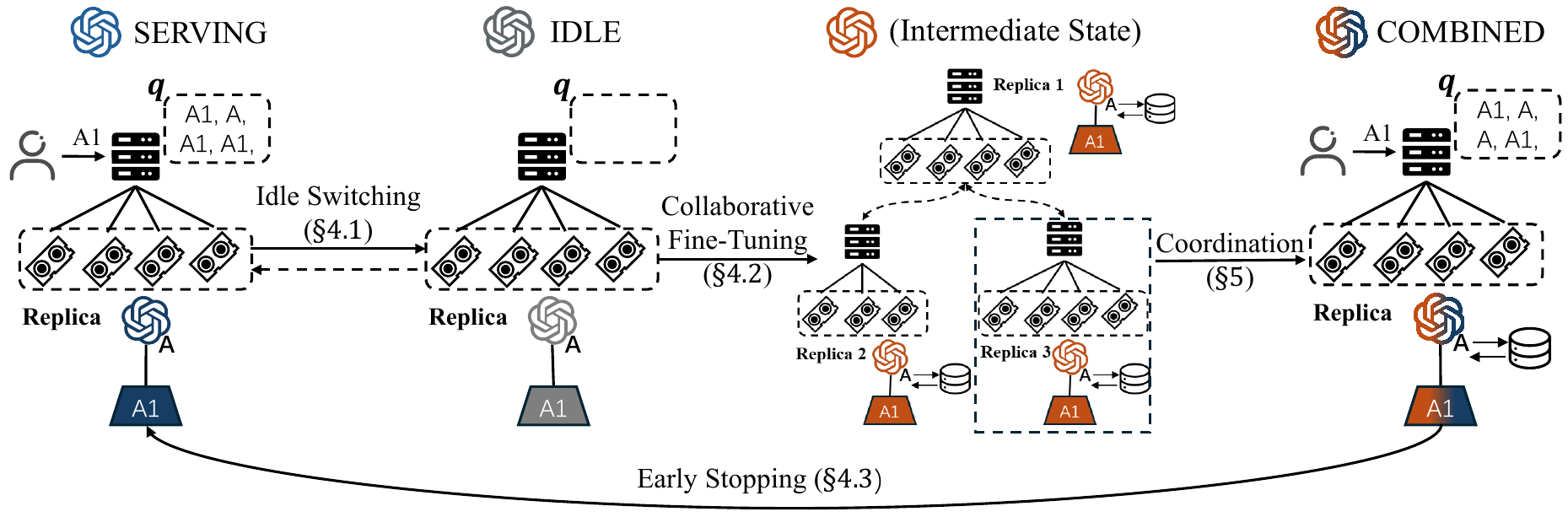}
    \caption{State switching in replicas.}
    \label{fig:states_all}
    \vspace{-0.3cm}
\end{figure*}

% \subsection{Replica's States in CoLLM}
% CoLLM为每个replica定义并维护三种虚拟（人为定义而非固有属性）的states: SERVING LIGHT SERVING，TRAINING and COMBINED。这些states界定了replica在CoLLM中所有可能出现的行为以及replica与CoLLM组件之间的交互。(具体解释干什么的)
\underline{\textbf{Replicas' States in CoLLM.}} As shown in Fig.~\ref{fig:states_all}, CoLLM maintains three system-defined states for each replica. These states characterize all the working behaviors of replica, as well as their interactions with system components:

\begin{itemize}[leftmargin=*]
\item \textsc{Serving}: The default operational state, in which the replica performs inference serving exclusively.
\item \textsc{Idle}: Represents replicas under sustained low workload, making them eligible candidates for fine-tuning.
\item \textsc{Combined}: Enables replica to concurrently perform long-running fine-tuning and real-time inference tasks on the same device and model.
\end{itemize}

\underline{\textbf{Inter-replica Co-orchestration.}} To utilize idle GPUs during low workload periods, CoLLM employs a dynamic co-orches-tration strategy driven by the Fine-tune Task Launcher, which enables flexible and on-demand initiation and stopping of fine-tuning tasks across eligible replicas.
% Co-orchestrating遵循了时间复用GPU资源的理念，首先。。。（简单介绍状态转移），然后Fine-tune Task Launcher。。。(简单介绍如何启动collaborative fine-tuning和早停)
Co-orchestration follows the principle of temporal multiplexing, where replicas alternate between inference and fine-tuning roles over time. Specifically, CoLLM detects replicas under sustained low workload and transitions them from \textsc{Serving} to \textsc{Idle}, making them candidates for fine-tuning. The Fine-tune Task Launcher then selectively initiates collaborative fine-tuning rounds with these candidates, while early stopping replicas that offer limited fine-tuning benefits.

% By integrating \textsc{Idle}-state transitions, collaborative fine-tuning, and early stopping, Fine-tune Task Launcher achieves flexible orchestration of fine-tuning and inference, leading to improved GPU utilization, better model quality, and reduced serving disruptions.

% 除此之外，CoLLM可以灵活的决定一个replica whether to participate in FL during each round. In other words, if a replica 对于整体模型的训练不再有帮助, it can % 触发replica的微调早停并使其从下一轮FL fine-tuning中被排除，节省下不必要的资源消耗全力支持inference seringuntil selected for training in subsequent rounds. 

% In CoLLM, base LLMs with their adapters in replicas can be shared between \textit{inference} and \textit{fine-tuning} tasks without duplicate deployment. It is established by switching the replica mode.

\underline{\textbf{Intra-replica Coordination.}} 
CoLLM enables coordinated spatial multiplexing via the Inference-training Coordinator, which adaptively configures per-replica batch sizes for both fine-tuning and inference tasks, based on the optimal training goodput under task interferences. By jointly optimizing these batch sizes, CoLLM maximizes the efficiency of fine-tuning while opportunistically utilizing leftover resources to serve inference requests, achieving a fully cost-effective balance between model improvement and system responsiveness.
% While co-orchestration enables temporal multiplexing of GPU resources, LLM fine-tuning remains a heavy and long-running workload compared to real-time inference. Additionally, due to diminishing benefits in fine-tuning effectiveness over time, maintaining a fixed training load often leads to suboptimal GPU utilization.

% To address these challenges, CoLLM enables coordinated spatial multiplexing via the Inference-training Coordinator.
% which seamlessly switches fine-tuning replicas to the \textsc{Combined} state to execute fine-tuning and inference tasks concurrently with shared model weights.

\underline{\textbf{Request Dispatching.}} 
% As a serving system, CoLLM's optimization process critically depend on its ability to manage unpredictable inference workloads. To this end, 
To handle unpredictable workloads, CoLLM introduces a subflow-based request dispatching mechanism, which transforms stochastic request streams into temporally regularized subflows aligned with replica's processing envelope. This design enables CoLLM to consistently maximize inference throughput and service quality, even under highly dynamic load conditions.

\section{Fine-tune Task Launcher}\label{sec:FTL}

\subsection{\textsc{Idle} State of Replica} 
% 首先，我们提出了replica的light serving状态，从而replica可以向Fine-tune Task Launcer声明自己当前准备好接受fine-tune任务。 但同时，由serving向light serving状态的转换需要谨慎地考虑，由于较少的转换会导致后续fine-tune任务难以展开，而较多的转换则会加剧剩余replica的服务压力。

CoLLM defines an \textsc{Idle} state to mark replicas under sustained low workload as candidates for fine-tuning and expose them to Fine-tune Task Launcher. Once the replica switches to \textsc{Idle} state, CoLLM will no longer dispatch requests until it resumes \textsc{Serving} or enters \textsc{Combined} state. Therefore, transitions to \textsc{Idle} must be carefully controlled: overly conservative transitions delay fine-tuning opportunities, while aggressive ones risk overloading the remaining replicas. CoLLM highlights two key insights in designing this transition:

\textit{(a) Low utilization alone is insufficient.} A replica with low GPU utilization may still have a backlog of pending requests. Switching such a replica prematurely could exacerbate queuing latency and shift burden to other replicas.

\textit{(b) Short-term fluctuations can mislead.} LLM workloads often exhibit both long-term trends and short-term bursts \cite{Wu2024}. Reacting to transient dips may trigger unnecessary state switches, leading to reduced system responsiveness.

In CoLLM, this transition is governed by a carefully designed condition that balances the response to instantaneous changes with the long-term trend of the system workload. More formally, a replica $i$ switches from \textsc{Serving} to \textsc{Idle} if the following conditions are met:
\begin{align}
    \tilde{U}_i^t < U_i^{\text{switch}}~and~\tilde{q}_i^t < q_i^{\text{switch}}, \label{eq:state_light}
\end{align}where $\tilde{U}_i^t$ and $\tilde{q}_i^t$ are exponentially weighted moving average (EWMA) of GPU SM utilization and queue length over a sliding window of $T$ time steps:
\begin{align}
    \tilde{U}_i^t &= \sum_{t'=t-T+1}^{t} \omega_{t'} \cdot U_i^{t'}, \quad
    \tilde{q}_i^t = \sum_{t'=t-T+1}^{t} \omega_{t'} \cdot |q_i^{t'}|,
\end{align}where $U_i^{t'}$ and $|q_i^{t'}|$ denote the utilization and queue length at time $t'$, and $\omega_{t'} = \frac{e^{-\lambda (t - t')}}{\sum_{s=t-T+1}^{t} e^{-\lambda (t - s)}}$ is the time-decay weight controlled by parameter $\lambda$. The thresholds $U_i^{\text{switch}}$ and $q_i^{\text{switch}}$ in Eq. \ref{eq:state_light} are the $\alpha$-quantiles of the EWMA metrics among all replicas in the cluster $\mathcal{N}$:
\begin{align}
    U_i^{\text{switch}} &= \text{min}(Q_{\alpha}(\{\tilde{U}_j \mid j\in \mathcal{N} \}),~U^{\mathcal{L}}), \\
    q_i^{\text{switch}} &= Q_{\alpha}(\{\tilde{q}_j \mid j\in \mathcal{N}\}),
\end{align}where $U^{\mathcal{L}}$ is a constant utilization lower bound and is set to 0.25.
% here, we restrict replicas to the same base model $m_i$ to isolate differences in resource usage across model architectures, while allowing to consider different adapters to capture workload variations for different request types.
Replicas in the \textsc{Idle} state are transitioned back to \textsc{Serving} under two conditions. First, if a replica remains unselected by the Fine-tune Task Launcher for $T'$ consecutive decisions, it is reverted to the \textsc{Serving} state to avoid prolonged idleness. Second, under elevated request load, the Dispatcher may immediately promote \textsc{Idle} replicas back to \textsc{Serving} to accommodate incoming inference requests, ensuring timely service and resource responsiveness.

The tuning of sliding window size $T$ and status rollback window $T'$ is adapted to the volatility of inference workload. When the workload exhibits high short-term fluctuation, larger $T$ and $T'$ help smooth out transient noise and avoid unnecessary transitions. Conversely, under more stable conditions, smaller window sizes enable faster responsiveness without sacrificing decision accuracy.

% \textsc{Idle}状态的replica若在连续的T'次决策周期下没有被Fine-tun Task Launcher选中参与fine-tuning，则会恢复\textsc{Serving}状态；此外在请求负载升高时，Idle的replica会被立刻回退到\textsc{Serving}状态以继续接受新的请求，这一过程由Dispatcher所控制。

% While $T$ could be dynamically tuned to match shifting workload patterns, the periodicity of LLM-serving workloads can remain stable for tens of minutes or more \cite{Zhang2023}, allowing CoLLM to adopt a fixed $T$ over extended periods that simplifies implementation while retaining sufficient adaptability.

% 需要说明的是，在之后随着Dispatcher的引入，replica将被期望按照Ideal Serving Mode进行推理任务，$|q_i^t|$约等于0.此时判断条件1中$|q_i^t|$将被更改为等效的infer batch size $b_i^t<b_i^{switch}$.其中$b_i^{switch}$的计算逻辑与$q_i^{switch}$相同。

Note that with the introduction of Dispatcher~(\S\ref{sec:dispatcher}), replica are expected to perform inference tasks according to the \textit{Ideal Serving Mode}, were $|q_i^t|\approx0$. In this situation, Eq. \ref{eq:state_light} will be changed to $\tilde{U}_i^t < U_i^{\text{switch}}~and~\tilde{b}_i^t<b_i^{switch}$, where $b_i^{switch}$ is calculated with the same logic as $q_i^{switch}$.

% \subsubsection{Limitations}
% This design does not tolerate deviations within the time window \( T \), which may lead to delayed transitions if rare anomalies occur. However, this trade-off prioritizes stability over aggressive responsiveness, consistent with our system's overall goals.

\subsection{Collaborative Federated Fine-tuning}
% 服务状态天然地区分了可以被用于微调任务的replicas，从而大大减少了Fine-tune Task Launcher的决策空间。然而，想要在replica状态多样，模型性能异构的环境下组织微调任务，并充分利用离散的空闲资源，Fine-tune Task Launcher仍然需要回答以下问题：
% 1.replica队列中的请求如何处理？
% 2.怎样利用多个相同模型的replica的资源？
The \textsc{Idle} state naturally filters out replicas that are eligible for fine-tuning, significantly reducing the decision space for the Fine-Tune Task Launcher. However, incorporating fine-tuning into a dynamic and heterogeneous serving system still poses two critical challenges:

\textit{(a) Managing pending requests.} When an \textsc{Idle} replica swit-ches to fine-tuning, its request queue must be handled properly to avoid service disruption: \textit{Should pending requests be retained, or redistributed to other replicas?}
% This decision must balance serving latency, resource utilization, and the potential overhead of redistributing.
% An alternative is to intelligently discard low-priority or expired requests to free up resources. 

\textit{(b) Leveraging distributed resources.} CoLLM deploys multiple replicas with identical base models and adapters, where training data from different tenants is distributed across replicas and typically cannot be centralized due to privacy and regulatory constraints. Independent fine-tuning per replica may miss the data diversity and parallelisation benefits.

% Moreover, for privacy and regulatory reasons, request data often cannot be centralized.

Addressing these challenges is critical to efficiently integrating fine-tuning tasks into a serving-centric environment. We next describe how CoLLM addresses these challenges:

\textbf{Managing Pending Requests.} In contrast to prior approaches that compare estimated efficiency across replicas to decide whether to retain or migrate requests, CoLLM simplifies this decision by allowing selected \textsc{Idle} replicas to continue serving all in-queue requests without interruption. This is achieved by transitioning the replica directly into the \textsc{Combined} state via the Inference-Training Coordinator, which enables concurrent fine-tuning and inference on the same model. This co-location eliminates request migration and adapter reloading overhead, streamlining system behavior and improving overall efficiency.

\textbf{Utilizing Distributed Resource.}
% 作为一种PEFT技术，LoRA本身就为微调提供了准确且高效的方法，在一个LoRA as a service的服务系统中，当需要继续微调模型时，很自然地会想到延续LoRA的方式，这保证了微调过程准确率与效率的下界。
In 'LoRA as a Service' system, leveraging LoRA for PEFT tasks is a natural choice, as it ensures a baseline level of training efficiency. To fully leverage scattered computation resources and data, we draw inspiration from FL and propose organizing selected \textsc{Idle} replicas into a Collaborative FL PEFT process, which enables collaborative fine-tuning across dispersed replicas. 

% 为此，我们参考最近涌现的结合PEFT与FL的工作，将Fine-tune Task Launcher启动fine-tune task的过程设置为FL LoRA的形式。首先，Fine-tune Task Launcher在一众候选replica（部署了所需foundation model且$s_i=-1$）中选择一个AOM最大的replica（AOM相同则选择可用资源最多的）作为gloabl model并负责aggregation.其余的n个replica则被视作FL架构中的clients。

% Fine-tune Task Launcher会实时监控系统中的Idle replica数量，每当新增了Idle replica，其就会进行一次启动判断：Launcher会找到所有部署了同一类模型的Idle replicas集合\mathcal{K},若|\mathcal{K}|>=3，则会在K上启动federated fine-tuning。具体来说，Fine-tune Task Launcher 首先 selects a replica with the best model quality score （replica上模型的属性，初始化为1，并随着federated fine-tuning进行更新） as the server to provide the initial global model and be responsible for aggregation. The remaining $n$ replicas are treated as clients (denoted as $\mathcal{N}$).

The Fine-tune Task Launcher continuously monitors \textsc{Idle} replicas in the system. Upon the appearance of new \textsc{Idle} replicas, the Launcher evaluates whether to initiate a new federated fine-tuning round. Specifically, it identifies the set $\mathcal{K}$ of \textsc{Idle} replicas serving the same model. If $|\mathcal{K}| \geq 3$, federated fine-tuning is triggered across replicas in $\mathcal{K}$. Among them, the Launcher selects the replica with the highest model quality score $Q$\footnote{an internal metric initialized to 1 and updated over the course of fine-tuning} to act as the server responsible for global model initialization and aggregation. 

% The remaining $|\mathcal{K}|-1$ replicas form the client set $\mathcal{N}$ and participate in the training phase.

In the FL PEFT process, each client and the server have a base LLM (with weights $\mathbf{W}_{pre} \in \mathbb{R}^{d \times l}$) and the same type of adapter (as trainable matrix $\Delta \mathbf{W} \in \mathbb{R}^{d \times l}$) kept in GPU, where $d$ and $l$ are the weight dimensions. In fine-tuning process, the parameters of the original base LLM are frozen, and only the incorporated adapters are trained. There are two main operations in each FL PEFT round. 
First, the server broadcasts the global adapters parameters $\mathbf{B}_g \in \mathbb{R}^{d\times r}$ and $\mathbf{A}_g \in \mathbb{R}^{r\times l}$ to the participating clients at the beginning of each round., where $\mathbf{B}_g$ and $\mathbf{A}_g$ are the low-rank matrices with rank $r$ and $\mathbf{B}_g\mathbf{A}_g=\Delta \mathbf{W}$.

The second operation involves local training and global aggregation. In each federated round, client receives the adapters parameters $\mathbf{B}_g$ and $\mathbf{A}_g$ and begins local model training using its local dataset $\mathcal{D}$. The training objective is to fine-tune the adapters by minimizing a local loss function: $F(\mathbf{W}) = \frac{1}{|\mathcal{D}|} \sum_{\xi \in \mathcal{D}} \ell(\mathbf{W}, \xi),$ where $\ell(\mathbf{W}, \xi)$ is the loss for model $\mathbf{W}$ at data sample $\xi$. After completing local training, clients upload their updated LoRA parameters to the server. Upon receiving updates from all clients, the server aggregates these parameters to form an global LoRA as follows~\cite{fedavg}:
\begin{equation}
    \overline{\mathbf{B}}^{(t+1)} = \frac{1}{|\mathcal{K}|} \sum_{k \in \mathcal{K}} \mathbf{B}_k^{(t)}, \quad
\overline{\mathbf{A}}^{(t+1)} = \frac{1}{|\mathcal{K}|} \sum_{k \in \mathcal{K}} \mathbf{A}_k^{(t)},
\end{equation} This aggregated model is then broadcast to clients in the next round. We adopt this strategy for its simplicity and stability, but CoLLM also supports other alternative FL strategies to better adapt to heterogeneous resource or data distributions.

% , we leave such extensions to future work.

Upon completing each round, the Launcher updates each participating replica’s model quality score as follows:
\begin{equation}
Q^{(t)} = Q^{(t-1)} \cdot \frac{F^{(t-1)} - F^{(t)}}{F^{(t-1)}},
\end{equation}
where $F^{(t)}$ denotes the averaged training loss aggregated from all clients in round $t$. 

% This global update ensures consistency across replicas and avoids instability in model quality-aware decisions such as batch size adaptation in Dispatcher. 

% While this approach does not capture fine-grained, per-replica variation in loss progression, it offers a more stable signal for guiding dispatching decisions during ongoing serving operations.

% Federated Fine-Tuning参考标准的FedAvg~\cite{fedavg}使用同步平均聚合，即The server等待所有clinet完成本地训练，之后updates the global LoRA modules accordingly to 
% \[
% \overline{\mathbf{B}}^{(t+1)} = \sum_{k \in \mathcal{S}^{(t)}} \mathbf{B}_k^{(t)} / m,\quad
% \overline{\mathbf{A}}^{(t+1)} = \sum_{k \in \mathcal{S}^{(t)}} \mathbf{A}_k^{(t)} / m
% \] and sends back to the next set of selected clients for the next communication round.
% 选择这种方式是因为模型参数的更新过程较为简单稳定，可以最小化fine-tuning过程与其他最优化决策之间的互相影响。但我们可以为Fine-tune Task Launcher更换不同的FL策略以适配异构的资源与数据，但这不在本文的探讨范围内。在完成全局模型更新与下发后，我们通过平均loss来更新每个参与replica的model quality score Q^t = Q^{t-1} * ((F^(T-1) - F^T) / F^(T-1))。

% 此外说明Launcher会选择合适的replica并进行必要的fine-tuning初始化准备,例如通讯，加载训练数据集，此时replica处于一个Training的中间态，但值得注意的是这一中间态并没有实际的存在时间，因为这些replica同时会被后文所介绍的Inference-Training Coordinator设置为Combined状态并从一开始就并发地进行微调与推理进程。

\subsection{Early Stopping}
In CoLLM's FL PEFT design, fine-tuning proceeds in rounds. After each round, each replica independently decides whether to participate in the next round based on local training effectiveness. Inspired by early stopping techniques in DL~\cite{prechelt2002early} and FL~\cite{9076082}, the Fine-tune Task Launcher monitors each replica’s local loss after training. If no further reduction in its training loss is observed, the Launcher considers the global model sufficiently adapted to the replica’s local data. In this case, the launcher excludes the replica from the next round of FL fine-tuning and transitions it back to the \textsc{Serving} state, making it it exclusively available to serve inference requests.

% This mechanism improves training efficiency and responsiveness to workload changes while avoiding redundant updates to well-fitted replicas.

% With the introduction of FL, replicas in fine-tuning task gain the flexibility to decide whether to participate in FL during each round. In other words, if a replica opts not to participate, it can trigger a fine-tuning early stop and be excluded from the next round of FL refinement-training, thus fully supporting subsequent rounds of training.

% 我们参考DL中的早停（或其他FL）机制，当一个replica在经过本地训练后loss不再下降，则代表全局模型已经较好的拟合了该replica上的数据，这时Fine-task Launcher会将其从下一轮FL fine-tuning中移除，并恢复为Serving状态。

\section{Inference-training Coordinator}\label{sec:ITC}

Long-period fine-tuning task is prone to diminishing per-iteration benefits (e.g., loss reduction) over time \cite{federated_problems}, and this inefficiency can be further amplified by the data and resource heterogeneity in FL PEFT. Some replicas may possess abundant GPU capacity but lack informative training data, vice versa. Therefore, treating all replicas equally over different periods leads to training inefficiency and suboptimal resource utilization. To address this challenge, CoLLM introduces the Inference-training Coordinator, which dynamically aligns training workload with replica capability.

% To address these limitations, CoLLM introduces the Inference-Training Coordinator to  concurrent fine-tuning and inference tasks on replicas, realizing the spatial multiplexing of GPU resources and coordinating intra-replica configurations to balance model quality improvement and serving efficiency.

% a per-round batch size optimizer for \textsc{Combined} replicas. By jointly modeling and adapting training and inference batch sizes, the Coordinator maximizes training \textit{goodput} while leveraging leftover resources for inference—achieving a fine-grained balance between model quality improvement and serving efficiency.

% How to coordinate the fine-tuning process to maximize effectiveness while balancing fine-tuning contribution to push the GPU utilization to the limit.
% 值得注意的是，以上的过程固然实用，但忽视了两个关键问题，一是数据质量的不均衡，另一个是资源的异构性。例如有的replica可能拥有充分的GPU资源但分布其上的数据对训练来说价值索然，反之亦然；一视同仁的对待这样的南辕北辙势必会造成训练效率的低下以及资源浪费。
% Notably, the above process is certainly practical, but ignores the key challenges of the uneven quality of data and resources heterogeneity. For example, some replicas may have sufficient GPU resources, but the data distributed on it is of little contribution for training, and vice versa; treating such a disparity equally will inevitably result in training inefficiency and waste of resources. In the following we describe how this challenge is addressed through in the inference-training coordinator.

\begin{figure}[tbp]
    \centering
    \includegraphics[width=\columnwidth]{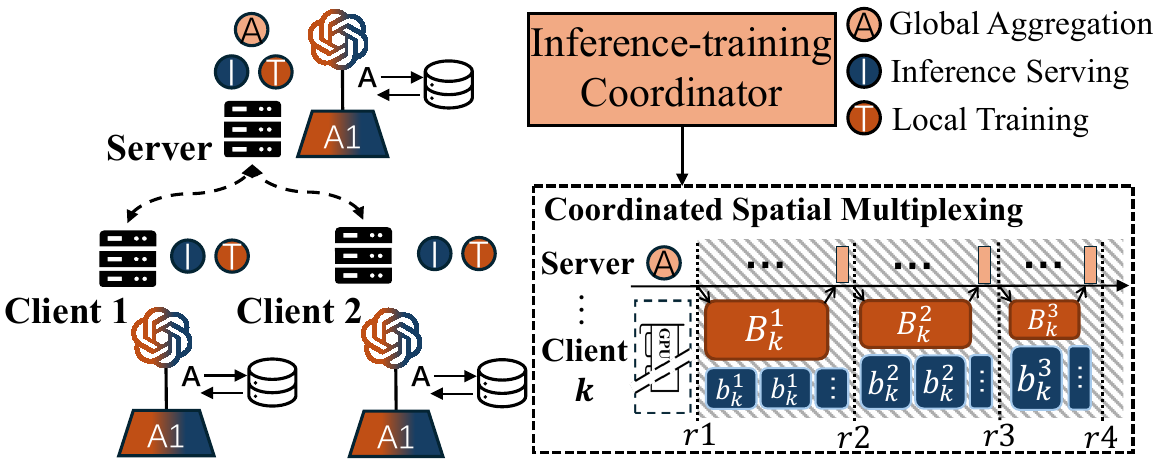}
    \caption{Inference-training coordinator.}
    \label{fig:IT_Coordinator}
    \vspace{-0.5cm}
\end{figure}

\subsection{Configuring Knobs and Objectives}

% CoLLM为每个由FT Launcher建立的Collaborative Fine-Tuning过程创建一个Coordinator，后者将审视所有参与的replica并将其状态切换为\textsc{Combined}，以允许replica以空间复用的方式并行地使用同一个模型进行fine-tuning和inference serving. then the coordinator manages the intra-replica configurations of the concurrent execution prcesses, aiming to maximize the effectiveness of local fine-tuning while maintaining SLO-compliant inference and high GPU utilization.
As shown in Fig.~\ref{fig:IT_Coordinator}, CoLLM creates a Coordinator for each FL PEFT process, which interrogates all participating replicas and switches their state to \textsc{Combined} to allow replicas to use the same model in a spatial multiplexing manner for fine-tuning and inference. On this basis, the Coordinator manages the intra-replica configuration of the co-located tasks, aiming to maximize the effectiveness of local training while maintaining SLO-compliant inference.

To quantify the effectiveness of fine-tuning, we adopt the concept of training \textit{goodput} from Pollux~\cite{Qiao2021}, which captures the trade-off between training throughput and convergence efficiency. In CoLLM, we extend this definition to account for the coexistence of inference and fine-tuning by introducing batch size configurations for both tasks. Specifically, for a \textsc{Combined} replica at round $t$ with training batch size $B_t$ and inference batch size $b_t$, denoted as $\mathcal{B}_t = (B_t, b_t)$, we define:
{\small
\begin{equation}
    \text{GOODPUT}_t(\mathcal{B}_t) = \text{THROUGHPUT}(\mathcal{B}_t) \times \text{EFFICIENCY}_t(B_t). \notag
\end{equation}
} \textbf{Throughput} is defined as the number of training samples processed per unit time:
\begin{equation}
    \text{THROUGHPUT}(\mathcal{B}_t) = \frac{B_t}{T_{train}(\mathcal{B}_t)},
\end{equation}
where $T_{train}(\mathcal{B}_t)$ is the per-iteration training time.\\
% , as discussed in \S 2, can be modeled as a linear function of both $B_t$ and $b_t$ to reflect interference between fine-tuning and inference tasks:
% \begin{equation}
%     y_1 = \alpha_1 B_t + \beta_1 b_t + \gamma_1
% \end{equation}
% \textbf{Efficiency-aware goodput.} The optimization objective integrates throughput and convergence efficiency, rewarding replicas that contribute meaningful training updates. Replicas with low efficiency (e.g., flat loss, high noise) receive reduced training batch sizes to release resources for inference.
\textbf{Efficiency} quantifies the training improvement per sample. Unlike Pollux, which considers only convergence speed, we introduce a weighted benefit model incorporating gradient noise $p_t$ \cite{McCandlish2018} and average per-iteration loss reduction $l_t$:
\begin{equation}
    \text{EFFICIENCY}_t(B_t) = \frac{a \cdot p_t \cdot l_t + B_0}{a \cdot p_t \cdot l_t + B_t},
\end{equation}
where $a$ is a scaling constant and $B_0$ is the initial training batch size. This formulation penalizes inefficient training as batch sizes grow and reflects diminishing marginal utility of additional compute. In addition to maximizing goodput, the Coordinator also configures inference batch size $b_t$ to be as large as possible while satisfies the latency SLO. 

% By jointly optimizing $(B_t, b_t)$, the Coordinator balances training throughput, convergence quality, and real-time serving demands.

% The inference latency $y_2$ is jointly influenced by both batch sizes and is modeled as:
% \begin{equation}
%     y_2 = \alpha_2 b_t + \beta_2 B_t + \gamma_2
% \end{equation}

% \begin{itemize}
%     \item The first priority is to optimize the efficiency of the fine-tuning process by maximizing its **goodput**, defined as:
%     \begin{equation}
%         \text{goodput} = \text{throughput} \times \text{efficiency}.
%     \end{equation}
%     \item Here, \textbf{throughput} is defined as the number of processed data samples per unit time, given by:
%     \begin{equation}
%         \text{throughput} = \frac{\text{batch size}}{\text{iteration time}}.
%     \end{equation}
%     \item The \textbf{training efficiency} measures the model performance improvement per processed data sample.
%     \item Furthermore, we aim to avoid GPU resource underutilization. Hence, the final optimization objective refines goodput by incorporating computational and memory resource efficiency:
%     \begin{equation}
%         \text{final objective} = \frac{\text{goodput}}{\text{resource usage (memory/SM)}}.
%     \end{equation}
% \end{itemize}

% \subsection{Decision Variables}
% \begin{itemize}
%     \item Computational resource allocation: $c_i^t$.
%     \item Memory allocation: $m_i^t$.
%     \item Batch size: $B_i^t$.
% \end{itemize}

\subsection{Interference-aware Coordination Algorithm}

The Inference-training Coordinator operates align with the FL PEFT round. Each round consists of three stages: initialization, performance modeling, and constrained optimization. In the first round, the Coordinator randomly assigns each replica a small training batch size $B_0$ and a relatively large inference batch size $b_0$. This conservative configuration mitigates early-stage overload, expedites draining of request queues, and allows the system to quickly collect representative metrics for performance modeling. Each replica trains for a small number of steps (50) to provide sufficient samples for the Coordinator to initiate model fitting.

After each round, the Coordinator collects runtime statistics including the average training iteration time $T_{train}$, training batch size $B$, gradient noise $p$, and average loss reduction $l$ for each participating \textsc{Combined} replica, along with inference latency $T_{infer}$ measured from served requests with inference batch size $b$. Using the throughput data collected so far, the Coordinator fits two separate latency models for fine-tuning and inference tasks.

% each as a function of $\mathcal{B}$, thereby explicitly capturing the interference arising from co-located execution of the two tasks on shared resources.

% \textbf{Modeling $T_{train}$ and $T_{infer}$.} % 正如2.2节所证明的，基于单一变量的线性建模无法精确建模训练和推理任务共享资源情况下的时延变化，为此我们的核心观点是通过多元线性函数同时拟合训练batch size B和推理batch size b，以此来捕获并发任务之间的干扰：
% \begin{align}
%     T_{train}(\mathcal{B}) = \alpha_{train} B + \beta_{train} b + \gamma_{train}, \\
%     T_{infer}(\mathcal{B}) = \alpha_{infer} b + \beta_{infer} B + \gamma_{infer}.
% \end{align} %这一建模并没有引入过多额外的开销，但却大幅提高了对并发任务的时延预测精度，尤其是当一种变量（如b）相同而另一变量（B）不同设置的情况下。下图展示了我们的T_{infer} model与一系列\mathcal{B}设置下的测量average inference time的拟合。总之，我们发现，T_{infer}在改变训练batch size和inference batch size时，都can represent the observed data closely

As demonstrated in \S\ref{sec:motivation2}, latency under concurrent fine-tuning and inference cannot be accurately modeled using single-variable linear functions. To capture the interference between the two tasks, CoLLM adopts a bivariate linear model that jointly considers the impact of both training batch size $B$ and inference batch size $b$:
\begin{align}
T_{train}(\mathcal{B}) &= \alpha_{\text{train}} B + \beta_{\text{train}} b + \gamma_{\text{train}}, \\
T_{infer}(\mathcal{B}) &= \alpha_{\text{infer}} b + \beta_{\text{infer}} B + \gamma_{\text{infer}}.\label{eq:latency_inter}
\end{align}

Compared to conventional latency models, our approach introduces minimal overhead but significantly improves
\begin{wrapfigure}{r}{0.55\linewidth}
    \centering
    \includegraphics[width=\linewidth]{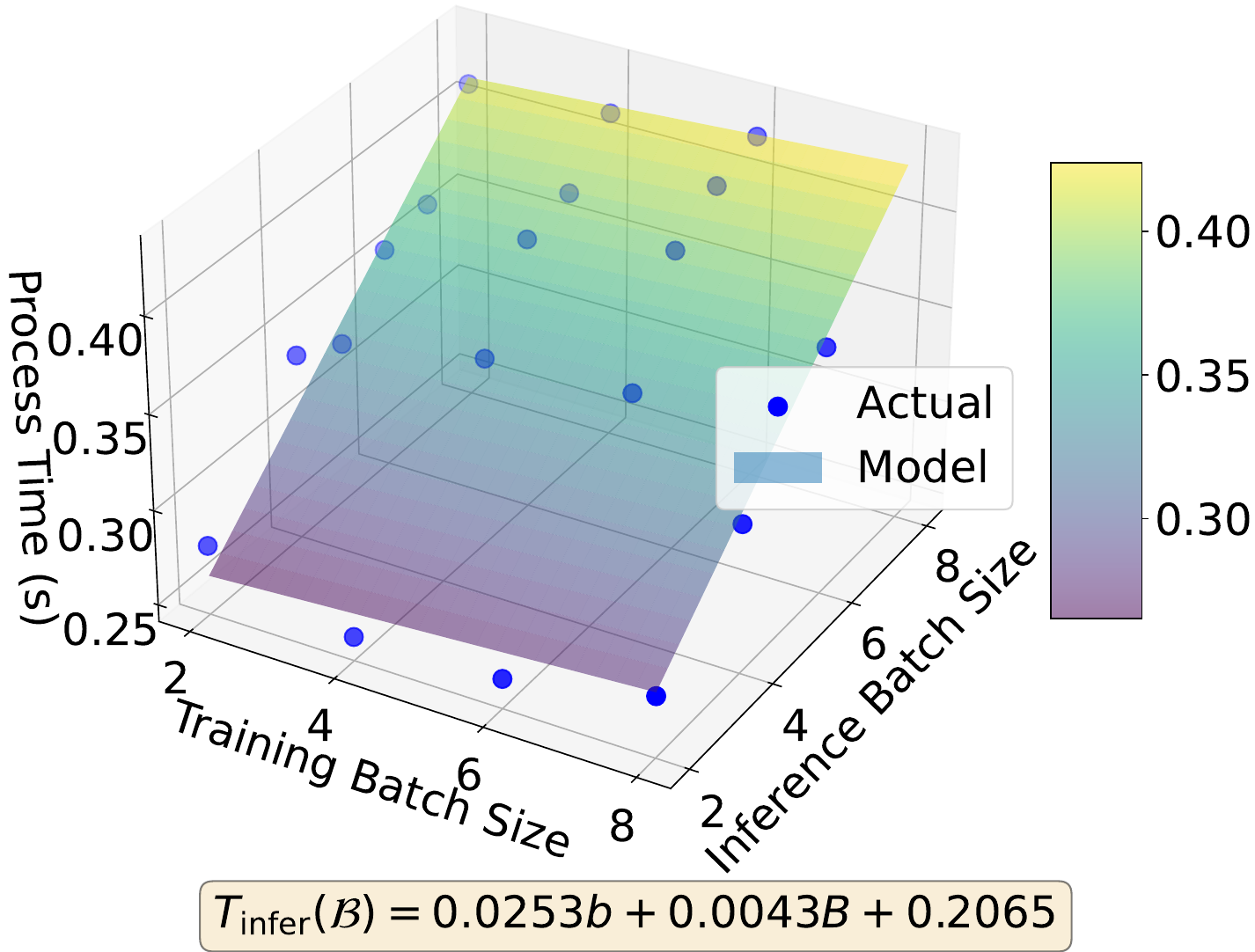}
    \caption{Validation of $T_{infer}$.}
    \label{fig:latency_modeling_val}
    \vspace{-0.3cm}
\end{wrapfigure} latency prediction accuracy under resource contention, particularly when one batch size remains constant while the other varies. As shown in Fig.~\ref{fig:latency_modeling_val}, the fitted $T_{infer}$ model closely matches the observed latencies acr-oss diverse batch configurations $\mathcal{B}$, effectively capturing task interference dynamics.

Given the fitted latency models, the Coordinator solves the following constrained optimization problem for each replica to determine the optimal fine-tuning and inference batch sizes $(B^*, b^*)$:
\begin{align}\label{eq:coordinator}
(B^*, b^*) &= \text{argmax}_{B} \ \text{GOODPUT}(B, b^{\star}(B)),
\end{align}
% where $\text{GOODPUT}(B, b)$ is expanded as:
% \begin{align}
%     \text{GOODPUT}(\mathcal{B}) &= \frac{B}{T_{train}(\mathcal{B})} \cdot \frac{a p l + B_0}{a p l + B_t},
% \end{align}
% with $B_0$ denoting the initial training batch size, and the two terms representing training throughput and training efficiency, respectively.
where $b^{\star}(B_t)$ denotes the maximum inference batch size satisfying the latency constraint $\tau'$ for a given $B$:
\begin{align}
b^{\star}(B) = \max\{b \ | \ \alpha_{infer} b + \beta_{infer} B + \gamma_{infer} \leq \tau' \},
\end{align}
and $\tau'$ is the process latency derived from the current request SLO and system queueing latency. Specifically, for each candidate $B$, the coordinator first computes the maximum feasible $b^{\star}(B)$ that ensures latency SLO compliance, then evaluates the corresponding $\text{GOODPUT}(B, b^{\star}(B))$, and selects the $(B^*, b^*)$ that maximizes the objective.

Once the optimal batch sizes are determined, the Coordinator propagates the updated configurations to the replicas for the next round. Simultaneously, the inference latency model parameters and optimized inference batch sizes are send to the Request Dispatcher to guide subflow pacing.

\section{Request Dispatcher}\label{sec:dispatcher}
% We first formulate the quality-aware dispatching problem and then present the subflow based dispatching algorithm to maximize inference throughput and service quality under highly dynamic load conditions.

\subsection{Problem Definition}

The goal of CoLLM's Request Dispatcher is to maximize system-wide inference \textit{Q-goodput}, defined as the number of requests that meet their SLO deadlines per unit time (goodput~\cite{Zhang2023}), weighted by request responses quality. To formalize this objective, we consider an idealized offline problem where the Dispatcher has access to the complete future.
% Let $\mathcal{R}$ be the set of all requests, where each request $r \in \mathcal{R}$ is characterized by its arrival time $a_r$, deadline $d_r$, and associated performance score $\text{perf}_r$. A batch $B$ is a group of requests scheduled together on a replica, with arrival time $a(B) = \max_{r \in B} a_r$ and deadline $d(B) = \min_{r \in B} d_r$.

% We define the offline scheduling objective as:
% \begin{align}
%     \mathbf{maximize} \quad & \sum_{t} \sum_{B \in \mathcal{B}_t} |B| \cdot \bar{\text{perf}}(B) \cdot \mathbb{1}\left[t + \ell(B) \leq d(B)\right] \label{eq:offline_obj}
% \end{align}
% where $\mathcal{B}_t$ is the set of batches scheduled at time $t$, $\ell(B)$ is the model-dependent latency of executing batch $B$, and $\bar{\text{perf}}(B)$ is the average performance score of all requests in $B$. The indicator function ensures that only requests served before their deadlines contribute to goodput.
% We define the offline serving problem as follows. 

Referring to \cite{Zhang2023}, we treat requests querying the same model and having the same latency SLO as an original request stream. 
Let $\mathcal{R}$ be the set of all inference requests in a stream, where each request $r \in \mathcal{R}$ is defined by its arrival time $a_r$, deadline $d_r$ and response quality score $Q_i(r)$ when served by replica $i$. Let $\mathbb{B}$ be the set of all valid batches, where each batch $\mathbf{B} \subseteq \mathcal{R}$ can be scheduled to a replica $i$ at time $t$; for a batch $\mathbf{B}$, arrival time $a(\mathbf{B})$ is the arrival time of the most recent request in $\mathbf{B}$, and deadline $d(\mathbf{B})$ is the earliest deadline of all requests in $\mathbf{B}$.  $x_{\mathbf{B},t,i} \in \{0,1\}$ denote whether batch $\mathbf{B}$ is assigned to replica $i$ at time $t$. Each batch $\mathbf{B}$ has a latency $\ell_i(\mathbf{B})$, and an average response quality score $Q_i(\mathbf{B})=\frac{1}{|\mathbf{B}|}\sum_{r\in \mathbf{B}} Q_i(r)$ when served by replica $i$. The offline optimization objective is to maximize the total Q-goodput:
\begin{align}
    \mathbf{maximize} \quad & \sum_{t} \sum_{i} \sum_{\mathbf{B} \in \mathbb{B}} x_{\mathbf{B},t,i} \cdot |\mathbf{B}| \cdot Q_i(\mathbf{B}) \label{eq:offline_obj} \\
   \text{s.t.} \quad & \sum_{t} \sum_{i} \sum_{\{\mathbf{B}|r\in \mathbf{B}\}} x_{\mathbf{B},t,i} \leq 1, \quad \forall r \in \mathcal{R}& \tag{a} \\
   & a(\mathbf{B})\cdot x_{\mathbf{B},t,i} \leq t, \quad \forall \mathbf{B},t,i& \tag{b} \\
   & (t + \ell_i(\mathbf{B}))\cdot x_{\mathbf{B},t,i} \leq d(\mathbf{B})), \quad \forall \mathbf{B},t,i& \tag{c} \\
   & \sum_{\mathbf{B} \in \mathbb{\mathbf{B}}} x_{\mathbf{B},t,i} \leq 1, \quad \forall t, i& \tag{d}
\end{align}
Subject to the following constraints:
\begin{itemize}[leftmargin=*]
\item[(a)] Each request is dispatched at most once.
\item[(b)] No request can be served before its arrival.
\item[(c)] Batches that miss their earliest deadline do not contribute to the objective.
\item[(d)] A replica executes at most one batch at a time.
\end{itemize}

This formulation assumes that batches can be optimally formed and scheduled across time. However, solving Eq.~\ref{eq:offline_obj} is unrealistic for two key reasons. First, request arrival patterns are highly unpredictable at sub-second granularity, making future-aware scheduling infeasible. Second, even with perfect foresight, selecting the optimal batching and scheduling decisions across replicas is NP-hard as it is a Zero-one Integer Linear Program (ZILP)~\cite{1164450}.

%  Due to these limitations, CoLLM adopts a latency modeling-based online dispatching algorithm that regulates request stream through subflows shaped to match each replica's ideal service mode (\S\ref{motivation3}), and adjusting their pacing adaptively to workload and model quality dynamics.

% 然而 (\S\ref{motivation3})的分析揭示了replica达到理论最大goodput的服务形态，即ideal service mode。这给我们设计CoLLM的Dispatcher提供了灵感，如果我们能regulates request stream to match each replica's ideal service mode, and adjusting their pacing adaptively to workload and model quality dynamics，则可以逼近Eq.~\ref{eq:offline_obj}的目标。

However, the analysis in \S\ref{motivation3} highlights a theoretical optimal serving pattern: \textit{Ideal Serving Mode}—in which each replica achieves its maximum possible goodput. This observation motivates our design of CoLLM’s Dispatcher. By regulating the request stream to align with each replica’s ideal serving mode and adaptively adjusting its pacing according to workload and model quality dynamics, we can approximate the optimal Q-goodput objective formalized in Eq.~\ref{eq:offline_obj}.

\subsection{Online Subflow-based Dispatching}
% Dispatcher的作用是将不可预测的随机请求流分流为具有固定特征，契合replica需求的子流，间接实现可预测性。参考shepherd。

% In large-scale LLM serving systems, inference requests typically arrive as stochastic streams, exhibiting diverse arrival rates, content lengths, and temporal bursts. Traditional dispatching schemes, such as round-robin scheduling or advanced per-request real-time assignment \cite{Wu2024}, often assume instant availability of processing capacity and perform on-the-fly decisions, which sacrifices batching efficiency and complicates latency guarantees under high-load conditions.

The key idea of CoLLM Dispatcher is to transform the unpredictable, fluctuating original request stream into a set of parallel subflows with tuned characteristics. Each subflow is mapped to a specific replica and configured to dispatch batched requests that match the replica’s ideal service mode. This design provides controllable request patterns to replicas, thereby indirectly achieving system-wide predictability and improved goodput under SLO constraints.

\paragraph{Subflow Execution.} For an original request stream, the Dispatcher maintains a subflow for each serviceable replica (with the queried model deployed). Each subflow $i$ is implemented as a thread that periodically samples $b_i$~\footnote{We use $b_i$  as it also determines the inference batch size on replica $i$.} requests from the original stream every interval $I_i$ and dispatches them to the assigned replica. Note that the actually fetched requests $b_{i,actual}$ do not always reach $b_i$, as the original request stream may be fetched empty $(b_{i,actual}\leq b_i)$. The sampling size and interval are determined  by the subflow's configuration, which is governed by the Dispatcher through a two-phase schedule: a \textit{macro-cycle} ($\mathbf{T}_{fit}$) for global latency model fitting and a \textit{micro-cycle} ($\mathbf{T}_{adjust}$) for localized subflow tuning. The design distinguishes between \textsc{Serving} and \textsc{Combined} replicas and applies different update rules.

\paragraph{Macro-Cycle ($\mathbf{T}_{fit}$):} Model-driven Batch Size Bounding. Every $\mathbf{T}_{fit}$ seconds, the Dispatcher collects per-replica inference metrics $(b, T_{infer}, \ell)$ from all \textsc{Serving} replicas, where the total latency $\ell$ is equal to the inference latency $T_{infer}$ plus the request queuing latency $T_{queue}$, and fits the exclusive inference latency model:
\begin{equation}\label{eq:latency}
    T_{infer}(b) = \alpha b + \beta.
\end{equation}
Based on the current average queuing latency $\overline{T}_{queue}$, we can derive the available execution budget for new requests:
\begin{equation}
\tau' = \tau - \overline{T}_{queue},
\end{equation}
where $\tau$ is the request latency SLO. The maximum feasible batch size without violating SLO constraints is calculated as:
\begin{equation}
b_{\max} = \left\lfloor \frac{\tau' - \beta}{\alpha} \right\rfloor.
\end{equation}
The Dispatcher then uses $b_{\max}$ to update the upper bound of each subflow’s batch size during the next micro-cycle. The latency model in Eq.~\ref{eq:latency} is also used to configure the subflow interval $I$ as $I=\alpha b_{actual}+\beta$. For \textsc{Combined} replicas, the maximum batch size $b_{max}$ is set as $b^*$,  which is determined by the Coordinator in Eq.~\ref{eq:coordinator}, and the interval is computed using the bivariate linear model in Eq.~\ref{eq:latency_inter}.

\textbf{Overload Mitigation.} If $\overline{T}_{queue} \geq \tau-\beta$, indicating inadequate service capacity and potential SLO violation, the Dispatcher proactively activates additional capacity by immediately promoting a serviceable \textsc{Idle} replica to the \textsc{Serving} state. After resource adjustment, macro-cycle fitting is re-executed with $\overline{T}_{queue}$ reset to $0.1\tau$ to update batch bounds.

% \begin{algorithm}[t]
% \caption{Micro-Cycle Adjustment}
% \label{alg:micro_cycle}
% \KwIn{Replica set $\mathcal{R}$, current batch sizes $\{b_i\}$, unsaturation rates $\{u_i\}$, model quality scores $\{Q_i\}$, max batch sizes $\{b_i^{\max}\}$, min batch size $b_{\min}$, smoothing ratio $\epsilon$}
% \KwOut{Updated batch sizes $\{b_i\}$}
% \vspace{0.5em}
% \textbf{Step 1: Compute Priority} \\
% \ForEach{$i \in \mathcal{R}$}{
%     $P_i \leftarrow Q_i \cdot (1 + u_i)$
% }
% Normalize $\{P_i\}$ so that $\sum_i P_i = 1$ \;
% \vspace{0.5em}
% \textbf{Step 2: Compute Ideal Allocation} \\
% $C_{\text{total}} \leftarrow \sum_{i \in \mathcal{R}} b_i^{\max}$ \;
% $C_{\text{remain}} \leftarrow C_{\text{total}} - |\mathcal{R}| \cdot b_{\min}$ \;
% \ForEach{$i \in \mathcal{R}$}{
%     $b_i^{\text{ideal}} \leftarrow b_{\min} + \lfloor P_i \cdot C_{\text{remain}} \rfloor$
% }
% \vspace{0.5em}
% \textbf{Step 3: Smoothed Update} \\
% \ForEach{$i \in \mathcal{R}$}{
%     $\delta \leftarrow \min\left( \epsilon \cdot b_i,~ |b_i^{\text{ideal}} - b_i| \right)$ \;
%     \uIf{$b_i^{\text{ideal}} > b_i$}{
%         $b_i \leftarrow \min(b_i + \delta,~ b_i^{\max})$
%     }
%     \Else{
%         $b_i \leftarrow \max(b_i - \delta,~ b_{\min})$
%     }
% }
% \end{algorithm}

\paragraph{Micro-Cycle ($\mathbf{T}_{adjust}$):} Quality-aware Subflow Adjustment. Every $\mathbf{T}_{adjust}$ seconds, the Dispatcher adjusts $b_i^t$ for each replica $i$ within the bound $b_{\max}$, guided by runtime feedback metrics: the unsaturation rate $u_i^t$ and the replica's model quality score $Q_i^t$. The $u_i^t$ reflects the degree to which subflow $i$ fails to fill its target batch size during the past interval:
\begin{equation}
    u_i^t = \text{mean}\left(\left\{ \frac{b_i^{t'}-b_{i,\text{actual}}^{t'}}{b_i^{t'}} \,\middle|\, t - \mathbf{T}_{adjust} \leq t' < t \right\}\right).
\end{equation}
Dispatcher then computes a priority score for each subflow:
\begin{equation}
    \text{Priority}_i^t = Q_i^t \times (1 + u_i^t).
\end{equation}which jointly favors replicas with higher model quality and lower batch saturation. The normalized priorities are used to allocate total batch capacity $\sum_j b_{max}$ to compute $b_i^t$:
\begin{equation}
    b_i^t = \sum_j b_{max} \cdot \frac{\text{Priority}_i^t}{\sum_j \text{Priority}_j^t}.
\end{equation}To prevent abrupt shifts, each $b_i^t$ is smoothed within the range $[\min(0.5b_i^{t-1}, 2), \max(1.5b_i^{t-1}, b_{\max})]$. This gradual adaptation ensures system stability while dynamically favoring more effective replicas.

% Batch sizes are redistributed proportionally to normalized priorities, ensuring that higher-quality, lower-unsatisfaction replicas are favored. Smoothing mechanisms prevent abrupt changes to avoid instability.

% For \textsc{Combined} replicas, only the subflow interval is updated (keeping the max batch size $b_{max}=b^*$ given by the Coordinator in Eq.~\ref{eq:coordinator}) to absorb requests during training without degrading fine-tuning performance.

\paragraph{The cost of controllability.} Compared to per-request dispatching, CoLLM’s subflow-based approach makes that replica handles inference requests at the ideal pace, avoiding idle cycles and backlog. 
However, it introduces queuing latency $T_{queue}$-requests may wait for their assigned subflow’s sampling window. This latency can increase end-to-end latency under low workload. However, this queuing overhead diminishes as the number of subflows increases, making it less pronounced in larger clusters.

\section{Implementation}\label{sec:implementation}

We implement CoLLM in Python with 4.2K lines of code. CoLLM uses the Hugging Face \texttt{transformers}~\cite{hf-transformers} and \texttt{PEFT}~\cite{peft} libraries to load base models and LoRA adapters into replicas, with \texttt{Accelerate}~\cite{accelerate} to deploy one LLM to multiple GPUs. CoLLM supports dynamical adapters binding where all adapters are loaded into CPU memory and swapped onto the GPU asynchronously when needed. However, adapter-level scheduling is not a focus of CoLLM as it has been explored at related works e.g.,  dLoRA~\cite{Wu2024} and S-LoRA~\cite{SLoRA}.

\textbf{Concurrent Inference and Fine-tuning via Model Sharing.} As stated in \S\ref{sec:model_sharing}, CoLLM supports concurrent fine-tuning and inference on a shared model within a single replica via Python’s \texttt{multiprocessing}. To enable this, we modify the Hugging Face \texttt{Trainer} class to support controlled training runtime. The default \texttt{Trainer} uses \texttt{Accelerate} for gradient states tracking and management, which performs strict model state check and assumes exclusive training ownership, causing runtime errors when inference accesses the same parameters. We override \texttt{Trainer} to bypasses \texttt{Accelerate}, disables strict state checks, and invoke PyTorch’s \texttt{.backward()} directly for local gradient updates. This design enables safe parallel execution with relaxed CUDA stream handling across subprocesses.

% This design supports relaxed CUDA stream behavior and allows fine-tuning and inference to run safely in parallel via separate subprocesses.

% Each \textsc{Combined} replica spawns two separate subprocesses: one for handling inference subflows and the other for executing fine-tuning. We use Python’s \texttt{multiprocessing} module with shared CUDA contexts, allowing concurrent access to the same model instance across tasks. Gradients and optimizer state are kept isolated to the training process to avoid interference.

\textbf{Inference-Training Coordinator and GPU Monitoring.}
We implement a grid search optimizer to solve the constrained goodput maximization problem in Eq.~\ref{eq:offline_obj}. The Coordinator exports the selected batch sizes and latency models to both the Launcher and the Dispatcher through a shared memory file and synchronizes via a lightweight messaging layer built atop ZeroMQ. We deploy the Monitor as a long-running process on each device, using NVIDIA’s \texttt{nvidia-smi} interface and the \texttt{pynvml} library to collect per-replica GPU memory and utilization statistics.

% All components run within Docker containers with pinned CUDA devices to simulate multi-GPU clusters on a single machine. In a full deployment, each replica would be deployed to a physical GPU in a multi-node setup using Ray actors and NCCL for inter-replica communication during federated aggregation.

\section{Evaluation}

\subsection{Experiment Setup}

% \textbf{Testbed.} All experiments are conducted on a private cluster with 4 GPU servers that connected by RDMA and NV link, each equipped with 8×NVIDIA H20 (96GB) GPUs and 2×Intel Xeon Platinum 8480 processors. We emulate a cluster of 16 logical replicas, each mapped to 2 GPUs, representing a typical LLM inference serving deployment. All software components are implemented in Python using PyTorch 2.1 and HuggingFace Transformers 4.39, with multi-process execution handled via \texttt{torch.multiprocessing} and \texttt{nccl} for inter-replica communication.
All experiments are conducted on a private cluster with 4 GPU servers, each equipped with 8 NVIDIA H20 (96GB) GPUs and 2 Intel Xeon Platinum 8457C CPUs. GPUs within each server are interconnected via NVLink, while the servers themselves are connected via RDMA-capable fabric. Base on this, we build a cluster of 16 logical replicas, each mapped to 2 GPUs, representing a typical configuration for multi-tenant LLM inference serving. All software components are implemented in Python using PyTorch 2.5 and HuggingFace Transformers 4.48, with \texttt{NCCL} for distributed communication.

\begin{figure*}[tbp]
    \centering
    \includegraphics[width=\textwidth]{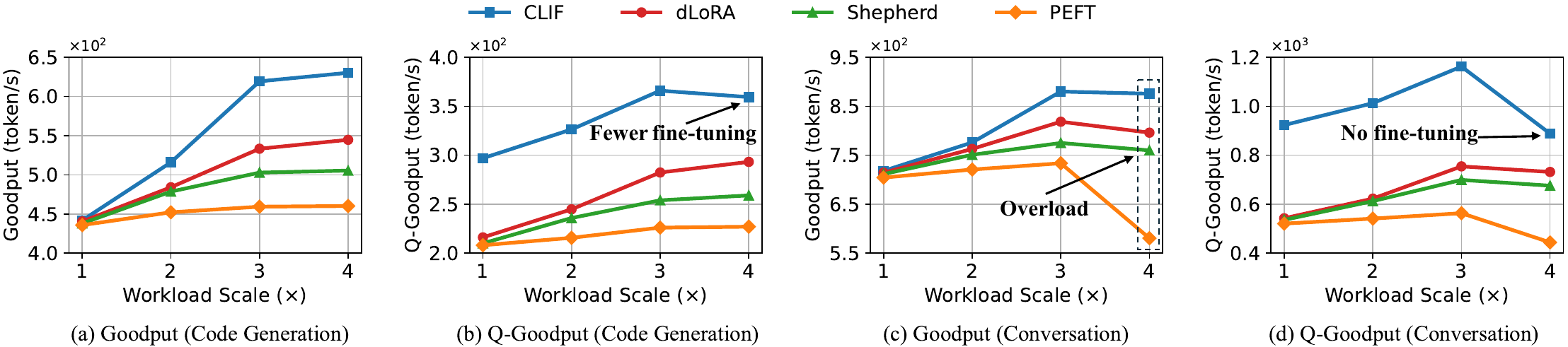}
    \caption{Goodput and Q-goodput of different serving systems with statistics for different task workloads separately.}
    \label{fig:throughput and goodput}
    \vspace{-0.5cm}
\end{figure*}

\begin{table}[tbp]
\caption{Inference and PEFT datasets.}
\label{table:datasets}
\begin{tabular}{lll}
\hline
\textbf{Dataset} & \textbf{Name}                    & \textbf{\# of Samples} \\ \hline
Code generation  & ManimCode~\cite{manim_code} & 4.4k \\
                 & CodeAlpaca~\cite{CodeAlpaca} & 20k \\ 
                 & CodeInstruct~\cite{code_instructions} & 122k \\\hline
Conversation     & alpaca~\cite{alpaca}                           & 50k                    \\
                 & GPTeacher~\cite{gpteacher}                        & 89.3k                  \\
                 & OpenInstruct~\cite{open}                 & 499k                   \\ 
                 & Instruct-3M~\cite{clean}                & 3.09M                  \\ \hline
\end{tabular}
\vspace{-0.5cm}
\end{table}

\textbf{Models and Initialization.}
% We evaluate CoLLM using three popular open-source LLMs with LoRA adapters: LLaMA-3.2-3B, LLaMA-3.1-8B, and LLaMA-2-13B (Qwen3-14B). The base models are loaded once into each replica’s GPU memory, and LoRA adapters are hot-swapped at runtime to support different tasks. For training, we use two PEFT datasets: (1) \textbf{Code}—merged from \texttt{iamtarun/code\_instructions\_120k\_alpaca} and \texttt{python\_code\_instructions\_18k}, covering diverse programming tasks, and (2) \textbf{Chat}—composed of \texttt{alpaca}, \texttt{Clean-Instruct-3M}, \texttt{GPTeacher-General-Instruct}, and \texttt{open-instruct-v1}, reflecting general instruction tuning. Each dataset is split 80/20 for fine-tuning and inference, respectively. For training, each replica receives a non-overlapping subset of 5k examples drawn uniformly at random from the corresponding training dataset.
 We evaluate CoLLM using two popular open-source LLMs: LLaMA-3.1-8B (16GB) and LLaMA-2-13B (26GB). All models are LoRA-instrumented and loaded once into GPU memory per replica. We summarize the collected LLM instruction-tuning datasets in Table~\ref{table:datasets}, each replica is preloaded with distinct datasets for fine-tuning to simulate a heterogeneous data distribution. The unloaded samples (about 30\%) are used as inference requests. (For input sequences, we use the ShareGPT [6] datasets for LLM inference and the Alpaca [53] dataset for PEFT, both are real-world datasets.)

\textbf{Workloads.}
To emulate realistic multi-tenant traffic, we replay the Azure-Code and Azure-Conv workloads~\cite{Azure_Dataset} described in \S\ref{sec:model_sharing} over a 10-hour window. The two traces are merged to simulate bursty, multi-tenant query patterns with different variations in arrival rate. In CoLLM, we evenly deploy different models and adapters to serve the merged workloads.

\textbf{Metrics.}
We focus on inference goodput, Q-goodput, and GPU utilization as our key metrics. Following prior work on LLM serving~\cite{Wu2024, sosp_23, yu2022orca}, we calculate goodput as the number of output tokens generated per second for responses meeting their SLO (0.5s per request~\cite{Wu2024}) and Q-goodput weights goodput by response quality, computed as the inverse of CE Loss. This token-level formulation extends request-based metrics to capture the true efficiency of LLM serving, where response lengths vary across prompts.

% along two token-level metrics: (1) Inference goodput~\cite{Zhang2023}: the number of output tokens generated per second for requests that meet their SLO deadlines. (SLO is set to 0.5s~\cite{Wu2024}); (2) Inference Q-goodput, computed as goodput weighted by response quality (the inverse of the CE Loss of the model responses); and (3) GPU utilization. 

% We also report latency to reflect end-to-end performance. 

% Overhead from system components (e.g., dispatching, coordination) is measured in wall-clock time per decision.

\textbf{Baselines.}
% We compare CoLLM with three state-of-the-art systems: (1) \textit{vLLM}~\cite{vllm}, a high-throughput inference engine with optimized KV cache management; (2) \textit{dLoRA}~\cite{Wu2024}, which supports dynamic adapter orchestration and fine-tuning-aware dispatching; (3) \textit{Shepherd}~\cite{Zhang2023}, which enables SLO-aware serving of DNN workloads with request scheduling optimizations; and (4) a vanilla \textit{PEFT}~\cite{peft} setup using HuggingFace's Transformers library. All baselines are re-implemented or ported to a unified cluster framework for fair comparison, and adapters are hot-swapped at runtime to support dLoRA and Shepherd's migration strategy.
We compare CoLLM with three state-of-the-art LLM inference serving systems: (1) \textit{dLoRA}~\cite{Wu2024}, which supports dynamic batching and requests migration; (2) \textit{Shepherd}~\cite{Zhang2023}, which enables SLO-aware serving of DNN workloads with optimized model assignment and batching; and (3) a vanilla \textit{PEFT}~\cite{peft} setup using HuggingFace's Transformers library. All baselines are ported to a unified cluster for fair comparison, and adapters are hot-swapped at runtime to support dLoRA and Shepherd's migration strategy.

\begin{figure}[tbp]
\centering
\subfloat[SM utilization]{\label{fig:sm_u}
\includegraphics[width=0.49\columnwidth]{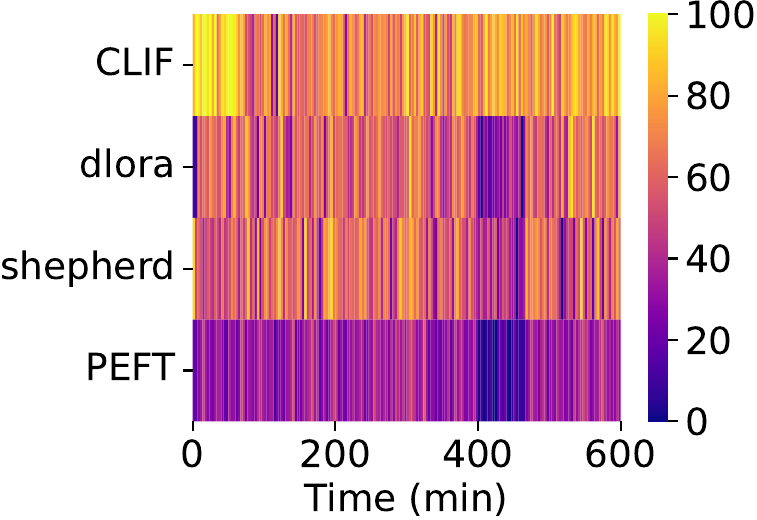}
}
\hspace{-0.27cm}
\subfloat[Memory utilization]{\label{fig:mem_u}
\includegraphics[width=0.49\columnwidth]{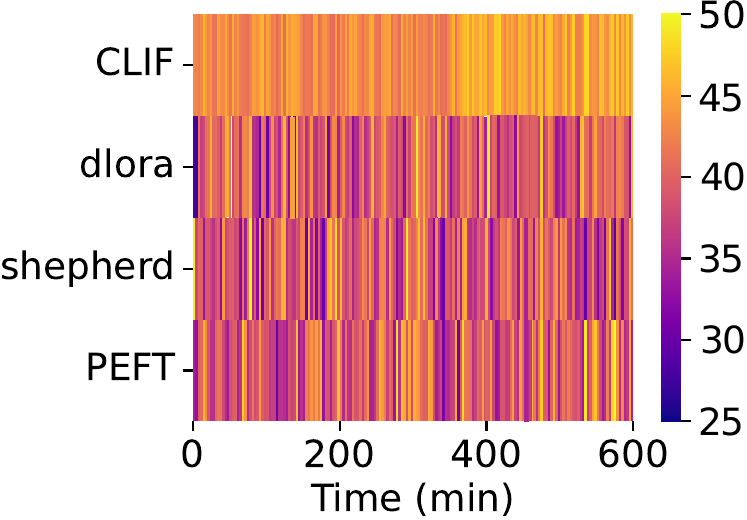}
}
\caption{Average GPU utilization in the cluster.}
\vspace{-0.5cm}
\end{figure}

\subsection{End-to-End Performance}

We first evaluate CoLLM's overall performance by comparing its goodput and Q-goodput against baselines under varying workload scales. To verify system scalability, we replay the workloads with four concurrency levels (1$\times$ to 4$\times$), while keeping the GPU budget fixed. Metrics are reported separately for conversation and code generation tasks.

\textbf{Inference Goodput.} As shown in Fig.~\ref{fig:throughput and goodput}a and Fig.~\ref{fig:throughput and goodput}c, under normal load scale (1$\times$), all systems can meet the SLO deadlines for most requests and exhibit similar goodput due to limited request volume and minimal batching opportunities. 
As workload scale increases, batching opportunities grow and goodput diverges. CoLLM consistently delivers the highest goodput across all scales and demonstrates the the most significant relative improvements over baselines at 4$\times$ workload. For the code generation workload, CoLLM delivers 1.16$\times$, 1.25$\times$, and 1.37$\times$ the goodput of dLoRA, Shepherd, and PEFT, respectively. 
Under the more demanding conversation workload, CoLLM maintains its advantage, achieving 1.1$\times$ the goodput of dLoRA, 1.15$\times$ that of Shepherd, and 1.51$\times$ that of PEFT.
These gains stem from CoLLM’s coordinated batching and proactive dispatching strategies, which enable consistent SLO-compliant serving and better resource saturation.

% compared to the reactive or static mechanisms employed in baseline systems.
% CoLLM’s subflow dispatching strategy scales favorably-the more replicas in the cluster, the less marginal queue latency per request, further amplifying batching benefits and goodput gains.

Among baselines, dLoRA benefits from adaptive request migration, while Shepherd prioritizes larger batches, both of which yield goodput gains over PEFT. Notably, the conversation workload sees higher goodput than code generation, owing to a larger volume of requests. However, as shown in Fig.~\ref{fig:throughput and goodput}c, at scale=4, all systems face overload and queuing of requests. In CoLLM, excess requests accumulate in the Dispatcher, while baselines queue within replicas. Although CoLLM mitigates this via controlled queue buffering and overload mitigation, persistent saturation still leads to SLO violations and reduced goodput across all systems.

\textbf{Q-Goodput.} As shown in Fig.~\ref{fig:throughput and goodput}b and Fig.~\ref{fig:throughput and goodput}d, CoLLM consistently achieves the highest Q-goodput across all workloads. Unlike baselines that serve with static models and thus exhibit constant response quality, CoLLM continuously improves model quality through fine-tuning, with updates immediately reflected in inference via model sharing.
% 由于其他系统并没有迭代模型的能力，其平均回复质量保持不变，同时CoLLM总能随着系统运行找到时机无缝地提升模型质量。
At low workload scale ($1\times$), where goodput differences are minimal, CoLLM outperforms baselines in Q-goodput by up to 1.68$\times$ (vs. PEFT on conversation). 
As the workload increases and inference dominates GPU usage, CoLLM maintains quality improvements through coordinated batch resizing and continuous fine-tuning in \textsc{Combined} mode, achieving up to 2.2$\times$ Q-goodput gains vs. PEFT at $3\times$ scale on conversation.
Notably, the Q-goodput gap between CoLLM and other systems narrows under extremely high load as training opportunities diminish. For instance, in the conversation workload at 4$\times$ scale (Fig.~\ref{fig:throughput and goodput}d), nearly all replicas are saturated, and CoLLM temporarily halts fine-tuning to prioritize inference. Consequently, Q-goodput decreases relative to 3$\times$ scale.
% Compared to conversation, the code generation task tends to yield higher Q-goodput due to lower CE loss per response—attributed to its more structured output format and stable decoding behavior. 

Overall, CoLLM significantly outperforms all baselines in both goodput and Q-goodput, scaling effectively while delivering high-quality, SLO-compliant service through seamless coupling of fine-tuning and inference.

% \textbf{Q-Goodput.} In terms of response quality, CoLLM consistently achieves the highest Q-goodput across all workload scales. Under light load (1$\times$), goodput differences are minimal, but CoLLM achieves up to 1.4$\times$ Q-goodput improvement by proactively launching fine-tuning on idle replicas, with model updates immediately reflected in inference via model sharing. As load increases, training opportunities diminish, and Q-goodput gains narrow, especially under scale=4 where fine-tuning becomes infeasible in the Conversation workload (Fig.~\ref{fig:throughput and goodput}(d)). This effect highlights CoLLM’s ability to opportunistically enhance model quality when resources permit, without harming serving throughput.

% Interestingly, Q-goodput in the Code Generation task is higher than Conversation, despite lower goodput, due to more structured outputs and lower per-token CE loss. Overall, these results confirm that CoLLM scales well in both throughput and response quality, delivering SLO-compliant, quality-aware serving even under dynamic and mixed workloads.

\textbf{GPU utilization.} To ecaluate resource efficiency, we measure GPU Streaming Multiprocessor (SM) utilization and memory utilization across the cluster under 3$\times$ workloads. Fig.\ref{fig:sm_u} shows the SM utilization temporal heatmap over the 10-hour period, averaged over all GPUs, while Fig.\ref{fig:mem_u} reports the corresponding memory utilization.

As illustrated in Fig.~\ref{fig:gpu_load}, CoLLM consistently sustains the highest SM utilization across time. This is primarily due to its ability to opportunistically orchestrate fine-tuning tasks on underutilized replicas, converting idle compute into training throughput. For instance, during a request dip between minutes 400–500, CoLLM maintains over 70\% utilization by activating fine-tuning workloads, while all baselines fall below 45\%. Moreover, CoLLM’s Dispatcher ensures that inference requests are dispatched in regular subflows that align with each replica’s ideal service mode, further boosting utilization. 
In contrast, dLoRA’s load-balancing strategy leads to better balanced utilization across time, but the underfilled batches may leave GPU compute cycles underused. Shepherd aggressively prioritizes larger batches and reassigns resources to high-load replicas, which results in higher variance—some GPUs are heavily loaded, while others idle. These strategies improve over the PEFT system, but still fall short under low-load conditions where batching alone becomes ineffective due to insufficient requests. CoLLM fills this gap by introducing auxiliary fine-tuning tasks to absorb spare residual capacity.

Regarding memory utilization (Fig.~\ref{fig:mem_u}), all baselines exhibit relatively similar memory footprints due to the dominant static memory overhead of LLM weights. CoLLM incurs slightly higher usage due to gradient and activation storage during fine-tuning, but LoRA’s lightweight design ensures memory remains within safe limits—preserving room for larger batches and multiple adapter.

% However, CoLLM achieves higher memory utilization, especially during training phases, as it retains gradient states and intermediate activations. Thanks to the use of LoRA-style PEFT, these additions incur minimal overhead and do not exhaust device memory, leaving headroom for large batches and multiple adapters.

\subsection{Microscopic Analysis}

%在这一章节，我们详细说明CoLLM的核心机制与组件的作用，为了公平对比，我们统一使用了scale=3$\times$的workload。
In this section, we examine CoLLM's core mechanisms in detail, and for a fair comparison, we uniformly use the workload under 3$\times$ scale.

% To examine the impact of CoLLM's model sharing design on response quality, we compare three configurations: (1) Model Sharing, where inference and fine-tuning co-execute on the same replica with shared model weights; (2) Separate, where fine-tuning and inference are executed independently and inference begins only after training completes; and (3) Inference Only, where no fine-tuning is applied.

% Fig.~\ref{fig:quality_cdf} shows the cumulative distribution (CDF) of response quality, measured as the inverse of the CE loss over a large set of evaluation prompts. CoLLM’s model sharing approach consistently achieves higher response quality than the other two baselines. Specifically, over 80\% of CoLLM responses exhibit quality scores above the 0.6 threshold, compared to only 63\% for Separate and 41\% for Inference Only.

% These gains stem from CoLLM’s ability to continuously incorporate fine-tuning improvements into the inference process in real time, without requiring model redeployment or disrupting ongoing requests. In contrast, the Separate approach introduces delays between training and serving phases, missing opportunities for timely adaptation. The Inference Only setting, while resource-efficient, fails to personalize models at all and thus produces the lowest-quality responses. This result confirms the effectiveness of CoLLM’s unified serving and fine-tuning model in delivering real-time, high-quality LLM responses.

\begin{figure}[tbp]
\centering
\subfloat[Response quality]{\label{fig:quality_cdf}
\includegraphics[width=0.48\columnwidth]{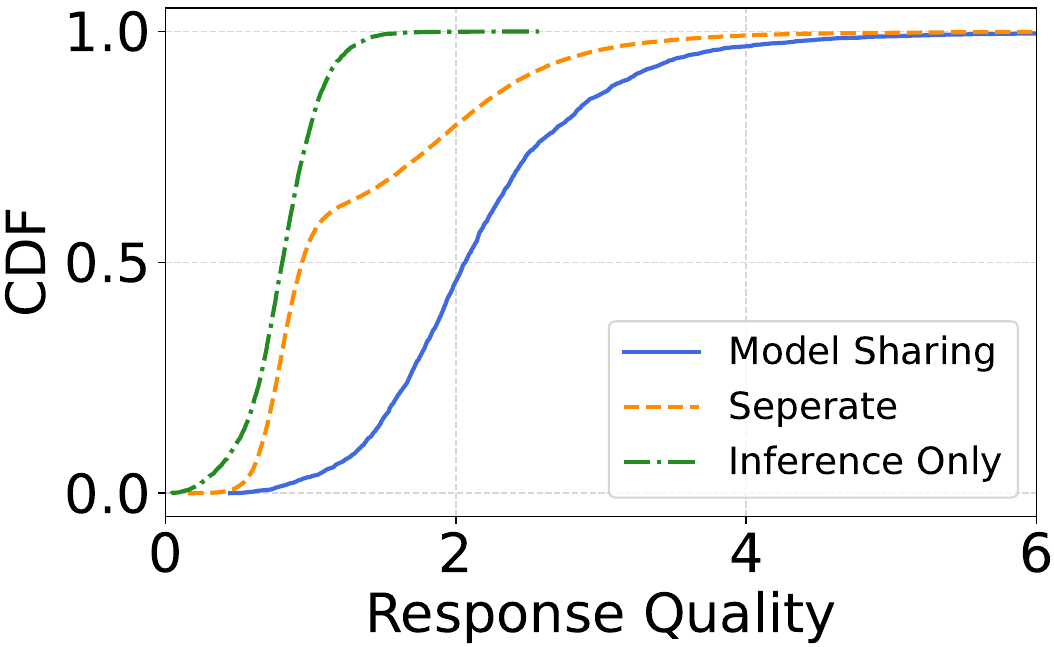}
}
\hspace{-0.1cm}
\subfloat[JCT and Q-Goodput]{\label{fig:coordinator}
\includegraphics[width=0.48\columnwidth]{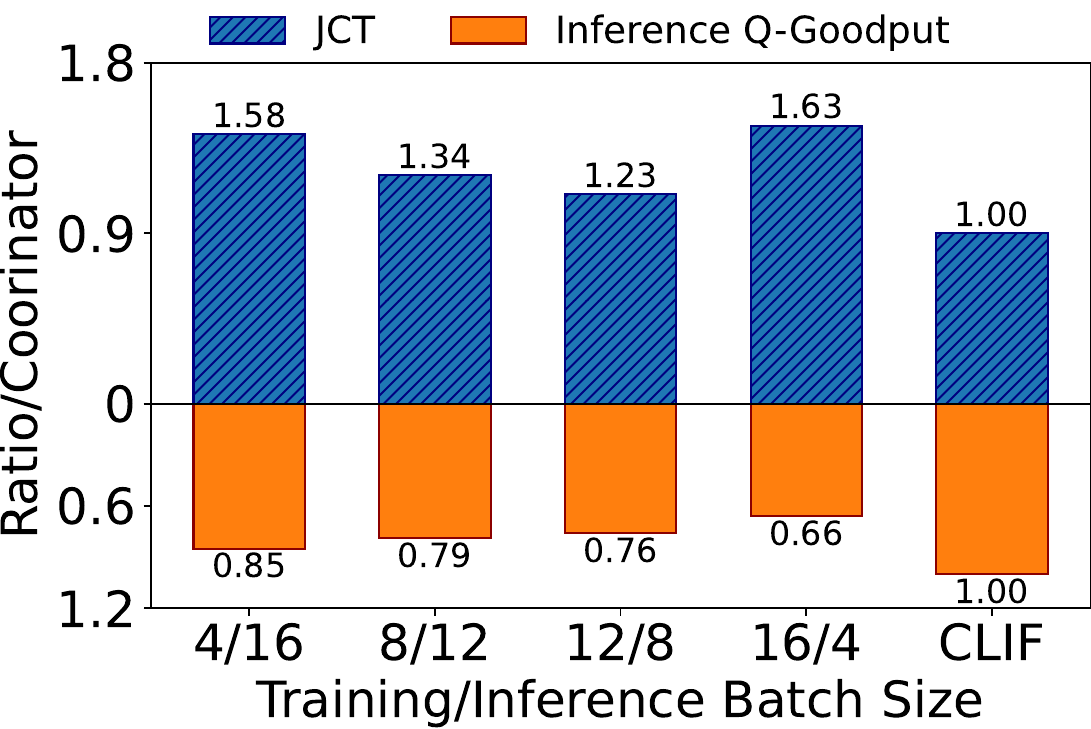}
}
\caption{The benefits of CoLLM's model sharing mechanism and Inference-training Coordinator.}
\vspace{-0.5cm}
\end{figure}

\textbf{Response quality improvement via model sharing.} To evaluate the effect of CoLLM’s model sharing design, we compare the response quality (inverse of the CE loss) of a \textsc{Combined} replica under three configurations: (1) Model Sharing, where inference and fine-tuning co-execute on the same replica with shared model weights; (2) Separate, where fine-tuning is performed independently and the updated model is only deployed after training completes; and (3) Inference Only. Fig.~\ref{fig:quality_cdf} shows the cumulative distribution of response quality, where both Model Sharing and Separate significantly outperform Inference Only. Notably, over 80\% of responses quality under Model Sharing exceed a quality score of 1, compared to fewer than 10\% for Inference Only.

However, a closer inspection reveals that CoLLM’s Model Sharing offers more consistent real-time gains. The Separate curve lags behind in the lower-quality region, e.g., there are still 60\% of Separate responses fall below a quality score of 1. This lag is due to the offline nature of the Separate—training improvements are not reflected in inference until model redeployment, leaving a substantial window with outdated model. In contrast, Model Sharing enables CoLLM to continuously feed training improvements back into inference, adapting on-the-fly to new requests, eliminating the cold-start penalty between training and serving phases. 

% These results validate CoLLM’s unified architecture as a principled approach to delivering timely, high-quality LLM responses.

\textbf{Dual goodput benefits of Coordinator.} We evaluate the effectiveness of CoLLM’s Inference-Training Coordinator by comparing five configurations for replicas in the \textsc{Combined} state: four with fixed training/inference batch sizes: (4,16), (8,12), (12,8), (16,4), and one enabled with Coordinator’s dynamic batch size adjustment. We use two metrics: inference Q-goodput and Joint Completion Time (JCT), defined as the time required for the federated model to reach a target quality level. Lower JCT indicates faster convergence and better training efficiency. All values are normalized to the Coordinator configuration for comparison.

From Fig.~\ref{fig:coordinator}, we can observe that the Coordinator achieves the best trade-off, yielding the highest Q-goodput while minimizing JCT. This outcome stems from its interference-aware latency modeling and dynamic goodput optimization, which jointly adjust per-replica batch sizes, adapting to diminishing training benefits and sustains dual-task efficiency.
In contrast, static configurations expose skewed focus. Settings like (8,12) and (12,8) favor training, achieving lower JCT but at the expense of inference Q-goodput. Meanwhile, increasing training batch size does not yield linear improvements—JCT under (16,4) is worse (JCT=1.63), suggesting that excessive batch sizes hurt training efficiency due to reduced update frequency and diminishing per-step gains.
On the other hand, (4,16) prioritizes inference and maximizes inference Q-goodput but significantly delays convergence (JCT 1.58), showing that overcommitting to inference leads to slow model quality improvement.

% Overall, CoLLM's Coordinator co-optimizes training efficiency and inference quality under hardware and SLO constraints, outperforming static configurations by dynamically adapting to workload patterns and fine-tuning returns.

\begin{figure}[tbp]
    \centering
    \includegraphics[width=\columnwidth]{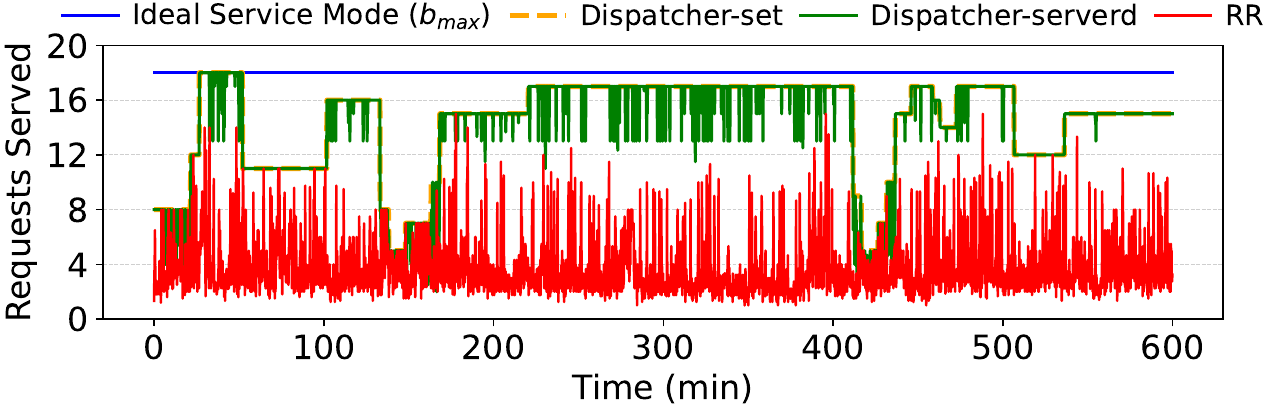}
    \caption{Stabilizing request serving via subflow-based dispatching.}
    \label{fig:dispatcher_performance}
    \vspace{-0.5cm}
\end{figure}

\textbf{Stability provided by Dispatcher.} To evaluate the effectiveness of CoLLM’s Dispatcher in improving goodput and stabilizing serving under dynamic workloads, we compare the real-time number of successfully served requests on a replica using CoLLM’s subflow-based Dispatcher (Dispatcher-served) versus round-robin dispatching (RR). The blue line shows the ideal service mode for the replica, while the Dispatcher-set indicates the batch size selected during micro-cycle adjustment. As shown in Fig.~\ref{fig:dispatcher_performance}, Dispatcher-set closely track the ideal service mode over time, with dynamic adjustments that reflect queue pressure and model quality shifts. The minimal gap between configured and actually served batches confirms the effectiveness of the Dispatcher in controlling the requests flow. In contrast, RR yields erratic and frequently degraded performance due to its uncontrolled dispatching. These results highlight that CoLLM’s subflow-based Dispatcher effectively regulates per-replica load, smoothing bursts and maintaining near-optimal serving under dynamic workloads.

% As shown in Fig.~\ref{fig:dispatcher_performance}, Dispatcher-set closely tracks the ideal service mode across time, and also dynamically adapt, with visibly oscillating to reflect queuing latency pressure and model quality changes. Moreover, the gap between the Dispatcher-set and actual Dispatcher-served remains minimal throughout, confirming that CoLLM successfully achieve controlled request pacing service. In contrast, the RR shows erratic served request and frequent dips, as it lacks load-aware adaptation and pushes requests indiscriminately—causing internal queuing delays and degraded throughput. These results demonstrate that CoLLM’s subflow-based Dispatcher effectively smooths load bursts and preserves serving efficiency by regulating request injection per replica, achieving near-ideal behavior under dynamic traffic conditions.

\begin{figure}[tbp]
\centering
\subfloat[Code Generation]{\label{fig:overhead_code}
\includegraphics[width=0.48\columnwidth]{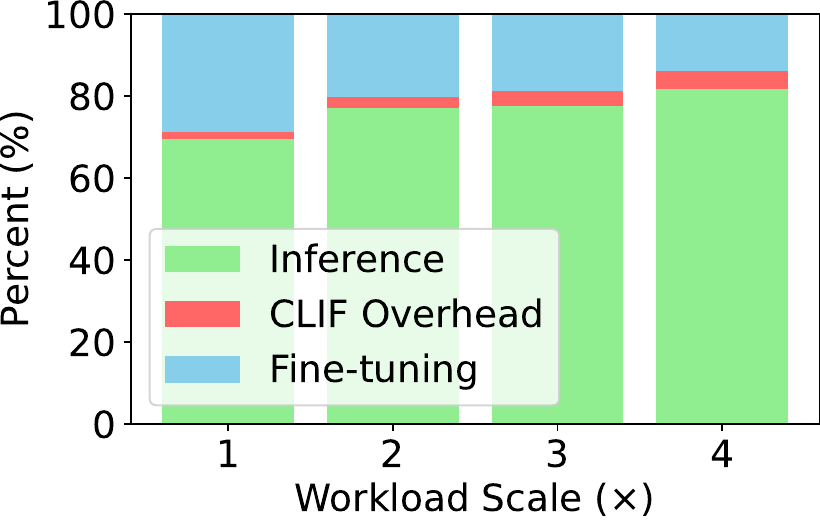}
}
\hspace{-0.2cm}
\subfloat[Conversation]{\label{fig:overhead_conv}
\includegraphics[width=0.48\columnwidth]{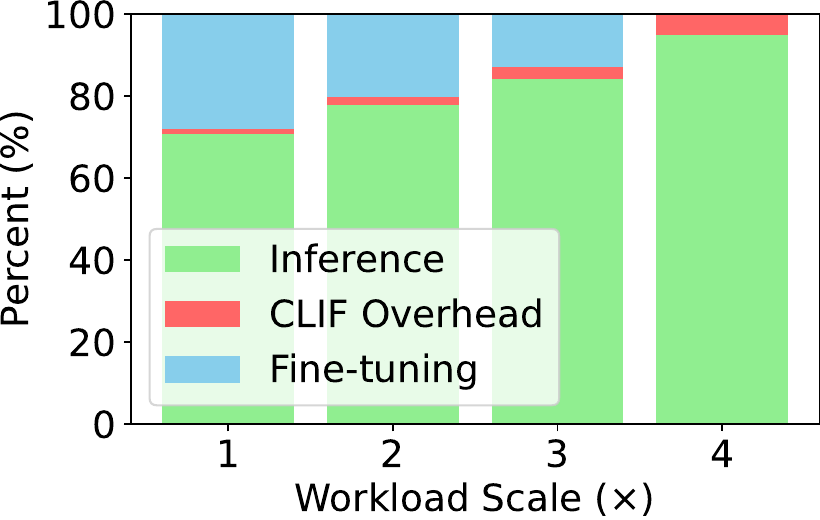}
}
\caption{CoLLM overhead under varying workload scales.}
\label{fig:CoLLM_overhead}
\vspace{-0.5cm}
\end{figure}

\subsection{System Overhead}
We evaluate CoLLM's runtime overhead by measuring the proportion of total processing time attributed to three categories: the time spent executing inference tasks (inference), the time spent on FL PEFT (fine-tuning), and CoLLM Overhead, which includes latency induced by the Dispatcher’s subflow pacing mechanism and the Coordinator’s optimization steps.
As shown in Fig.~\ref{fig:CoLLM_overhead}, CoLLM Overhead remains consistently low across both workloads, accounting for less than 2\% on average and never exceeding 5\%, even under 4$\times$ workload scale. This confirms that CoLLM’s coordination and dispatching logic incurs minimal overhead relative to serving and training workloads. We also observe that when the system operates under light load (scale=1), approximately 30\% of compute time is orchestrated to fine-tuning, enabling the system to opportunistically improve model quality. As the workload scale increases, fine-tuning time declines as resources are increasingly needed for inference. Under high load (e.g., 4$\times$ for conversation in Fig.~\ref{fig:overhead_conv}), inference dominates compute time and fine-tuning is entirely suspended. This adaptive orchestration validates CoLLM’s design: leveraging idle resources for fine-tuning without interfering with real-time serving.

% \begin{figure}[tbp]
%     \centering
%     \includegraphics[width=\columnwidth]{pics/Figure_2.png}
%     \caption{Comparison of fixed configuration.}
%     \label{fig:tmp_results}
% \end{figure}

\vspace{-0.5cm}
\section{Related Work}
% 我们在第二章详细介绍了与CoLLM设计思想相关的技术和工作，这一章将围绕现成的服务系统进行介绍。可以参考FlexMM
We detailed the mechanisms and techniques associated with the CoLLM design ideas in \S\ref{sec:motivation}, the following is organized around existing advanced DL service systems.

\textbf{DL serving systems.} Existing DL systems typically optimize training or inference in isolation. Pollux~\cite{Qiao2021} improves training efficiency via dynamic batch sizing while SiloD~\cite{silod} co-schedules compute and storage to reduce training JCT. On the inference side, dLoRA~\cite{Wu2024}, Shepherd~\cite{Zhang2023}, and ORCA~\cite{yu2022orca} enhance throughput and SLO compliance by dynamically adjusting batching, model placement, or request scheduling. However, these systems inevitably underutilize resources during off-peak periods, as the throughput bottleneck comes from workload scale rather than system capacity.

\textbf{Joint inference and training Systems.}
Recent work explores training–inference co-scheduling to improve cluster resource utilization. DeepBoot~\cite{Chen2023} and Lyra~\cite{lyra} opportunistically launch training on idle inference GPUs, while Mudi~\cite{mudi2024} adapts co-located tasks workloads via offline latency modeling. These systems treat training and inference as disjoint tasks, introducing costly model redeployment and missing opportunities to improve serving quality during runtime.

% and do not support real-time quality adaptation as they miss the potential of model sharing.
\vspace{-0.5cm}
\section{Conclusion}
This paper presents CoLLM, the first co-orchestrating system that unifies LoRA-based LLM inference and fine-tuning on shared replicas within GPU clusters. CoLLM improves resource utilization and inference goodput by enabling temporal and spatial GPU multiplexing through model-level sharing, leveraging idle resources and prioritizing quality-aware serving.
To achieve this, CoLLM dynamically orchestrates fine-tuning and inference via replica state management and FL PEFT integration, adapts intra-replica batch sizes through interference-aware coordination, and proactively regulates per-replica serving behavior with subflow-based dispatching.
Evaluation on production-scale traces show that CoLLM significantly improves both inference goodput and response quality, while maintaining high GPU utilization.
CoLLM holds strong potential for emerging personalized LLM service applications, such as AI assistants and autonomous agents, that require real-time model adaptation from continuous user queries.

\begin{acks}
To Robert, for the bagels and explaining CMYK and color spaces.
\end{acks}

\bibliographystyle{ACM-Reference-Format}
\bibliography{sample-base}

% \appendix









\end{document}